\documentclass{aa}  
\usepackage{graphicx}
\usepackage{txfonts}
\usepackage{lipsum}
\usepackage{subcaption}                                     
\usepackage{lscape}                                     
\usepackage{placeins}          
\usepackage{xcolor}                               

\usepackage{hyperref}

\begin{document}

   \title{Probing turbulence in the ionized interstellar medium using young pulsars}
   \subtitle{I. Dispersion measure variations}

   \author{S.~J.~I.~Melfor\inst{1,2}\corrauth{melfor@astron.nl}
   \and A.~Parthasarathy\inst{1,2,3}
   \and S.~Johnston\inst{4}
   \and A.~Chalumeau\inst{1,5}
   \and S.~Dai\inst{4}
   \and M.~Kerr\inst{6}
   \and M.~E.~Lower\inst{7}
   \and R.~N.~Manchester\inst{4}
   \and L.~S.~Oswald\inst{8,9,10}
   \and S.~Roch\inst{7,11,4}
   \and C.~A.~Sobey\inst{4}
   \and P.~Weltevrede\inst{12}
   }

   \institute{ASTRON, Netherlands Institute for Radio Astronomy, Oude Hoogeveensedĳk 4, 7991 PD Dwingeloo, The Netherlands
             \and 
             Anton Pannekoek Institute for Astronomy, University of Amsterdam, Science Park 904, 1098 XH Amsterdam, The Netherlands
             \and
             Max-Planck-Institut f\"{u}r Radioastronomie, Auf dem H\"{u}gel 69, 53121, Bonn, Germany
             \and
             Australia Telescope National Facility, CSIRO, Space and Astronomy, PO Box 76, Epping, NSW 1710, Australia
             \and
             LPC2E, OSUC, Univ Orleans, CNRS, CNES, Observatoire de Paris, F--45071 Orleans, France
             \and 
             Space Science Division, Naval Research Laboratory, Washington, DC 20375--5352, USA
             \and
             Centre for Astrophysics and Supercomputing, Swinburne University of Technology, PO Box 218, Hawthorn, VIC 3122, Australia
             \and
             Astrophysics Group, Cavendish Laboratory, University of Cambridge, J. J. Thomson Avenue, Cambridge CB3 0HE, UK
             \and
             School of Physics \& Astronomy, University of Southampton, Southampton SO17 1BJ, UK
             \and
             Department of Astrophysics, University of Oxford, Denys Wilkinson Building, Keble Road, Oxford OX1 3RH, UK
             \and
             OzGrav: The ARC Centre of Excellence for Gravitational-Wave Discovery, PO Box 218, Hawthorn, VIC 3122, Australia
             \and
             Jodrell Bank Centre for Astrophysics, Department of Physics and Astronomy, The University of Manchester, Manchester, M13 9PL, UK}
             
   \date{Received XXXX XX, 20XX}

\abstract
{Temporal variations in the dispersion measure (DM) of pulsars probe fluctuations in the free-electron density of the ionized interstellar medium (IISM). While DM variations have been studied extensively in millisecond pulsars, young pulsars provide a complementary probe due to their lines of sight (LoSs) intersecting supernova remnants, pulsar wind nebulae, and other turbulent environments. However, systematic studies of DM variability in young pulsars remain limited.}
{We characterize the temporal DM variations of a large population of young pulsars to constrain the turbulence of the IISM along their LoSs and to investigate the physical origins of the observed variability.}
{We analyzed 18 years of timing observations of 364 young pulsars obtained with the Murriyang/Parkes radio telescope, combining legacy data from the 21~cm Multibeam receiver with recent observations from the ultra-wideband low-frequency (UWL) receiver. Epoch-wise DM measurements were derived from sub-banded times of arrival, and the variability was characterized using linear trends and DM structure functions, which were compared with expectations for a Kolmogorov turbulence.}
{Significant DM variations are detected in 89 pulsars in the UWL data and in 23 pulsars in the multibeam data, with 16 in the combined dataset showing significant trends. This represents one of the largest long-baseline studies of DM variability in pulsars. We find tentative evidence that the observed DM variations are predominantly consistent with Kolmogorov turbulence. PSRs J1105--6107 and J1829--1751 show estimated DM variations driven by periodic profile variability. We also identify 21 bright, slow pulsars exhibiting short-term DM variations that exceed expectations from the IISM.}
{Long-baseline timing of young pulsars is a powerful probe of IISM turbulence on large scales. The emergence of DM measurements limited by intrinsic pulse-to-pulse variability highlights a key limitation for IISM studies at next-generation facilities with high sensitivity and large fractional bandwidths.}


\maketitle
\nolinenumbers

\section{Introduction}\label{sec:introduction}
The rotational stability of pulsars enables the high-precision timing of their beamed radio emission, which is observed as periodic pulses when the beam intersects our line of sight (LoS) to the pulsar. The turbulent ionized interstellar medium (IISM) imprints itself onto every radio pulse that traverses it, and among the most direct tracers of this turbulence are the temporal variations in the dispersion measure (DM). The DM is the integrated free electron density along the LoS to a pulsar,
\begin{equation}
    \text{DM} = \int_0^d n_e(l)\text{d}l,
    \label{eq:dm}
\end{equation}
where $d$ is the pulsar distance and $n_e$ is the free electron density along a LoS $l$. Because of the dispersive nature of free electrons, a radio pulse accumulates a frequency-dependent propagation delay as it travels through the IISM. This time delay is given by \citep{manchester1972}
\begin{equation}
    \Delta t_{\rm DM} \simeq \frac{1}{2.41 \times 10^{-4}} \times \text{DM} \times \left(\frac{1}{\nu_1^2} - \frac{1}{\nu_2^2} \right)\text{ s},
    \label{eq:t_dm}
\end{equation}
where $\nu_1$ and $\nu_2$ are two different observing frequencies. Because the dispersive delay scales as $\nu^{-2}$, multifrequency observations enable precise DM estimation, with lower frequencies providing the strongest leverage.\footnote{In practice, frequency-dependent profile evolution complicates the determination of an absolute DM value \citep[e.g.,][]{ahuja2007, hassall2012}.}

Consequently, temporal changes in the free electron content along the LoS manifest themselves as variations in the measured DM. As the pulsar, Earth, and IISM all move relative to one another, the LoS sweeps through different regions of the electron density distribution, and the resulting DM variations encode information about the structures encountered along the way \citep[e.g.,][]{rankin1971}. These variations take several forms. Linear trends can arise when the pulsar's transverse velocity dominates the relative motion, causing the LoS to drift steadily through a density gradient \citep[e.g.,][]{keith2013,lam2016}. 

Evidence of such variability dates back nearly five decades. \citet{isaacman1977} discovered DM variations in the Crab pulsar (PSR~B0531+21), where the DM increased by approximately 0.02~pc~cm$^{-3}$~yr$^{-1}$ over a six-year period as the LoS evolved within its host supernova remnant (SNR; \citealt{lyne1988}). Around the same time, the Vela pulsar (PSR~J0835--4510) was found to have a decreasing DM of 0.04~pc~cm$^{-3}$~yr$^{-1}$, which was attributed to the pulsar's transverse motion sampling different parts of the Vela SNR \citep{hamilton1977, hamilton1985}. 

Annual modulations can be introduced by the Earth's orbit, which changes the solar elongation angle and, therefore, the solar wind contribution to the DM \citep[e.g.,][]{you2007_solar, tiburzi2019, susarla2024}. In addition, high-density plasma structures, such as those responsible for extreme scattering events, can act as diverging lenses and produce localized transient DM excursions \citep[e.g.,][]{coles2015,kerr2018}. Superimposed on these trends are stochastic fluctuations driven by turbulence and anisotropy in the IISM (see Eq.~\ref{eq:spectrum}; e.g., \citealt{rickett2002,brisken2010}).

These early detections hinted at a broader picture in which the stochastic component of DM variations directly probes the spatial power spectrum of electron density fluctuations. A wide range of observations, from diffractive scintillation on scales of $\sim 10^6$--$10^9$~m to long-term DM monitoring on scales of $\sim 10^{12}$--$10^{14}$~m, indicate a power-law wavenumber spectrum for these fluctuations \citep[e.g.,][]{armstrong1995, cordes1998, stinebring2001},
\begin{equation}
    P_{\delta n_e}(q) = C_n^2\, q^{-\beta}, \quad
    2\pi/l_0 \leq q \leq 2\pi/l_i,
    \label{eq:spectrum}
\end{equation}
where $C_n^2$ is the spectral coefficient, $q$ is the spatial wavenumber, $\beta$ is the spectral index, and $l_0$ and $l_i$ are the outer and inner scales of the turbulence, respectively \citep{rickett1977, coles1987}. A Kolmogorov energy cascade predicts $\beta = 11/3$ \citep{rickett1990}. The value of $\beta$ determines the nature of the fluctuations in the DM time series: low values produce white-noise-like fluctuations, while high values of around $11/3$ or higher yield correlated, quasi-linear behavior. Observational constraints from pulse scattering restrict $\beta$ to the range $2 < \beta < 4$ for length scales $l_i \ll l \ll l_0$ \citep[e.g.,][]{lam2015}. By measuring $\beta$ through the observed DM variations, we can therefore diagnose the nature of the turbulence along a given LoS and identify environments where local structures such as SNRs and H\,{\sc ii} regions drive the fluctuation amplitude to above that expected for the diffuse IISM.

The connection between DM variability and the pulsar's environment has been exploited in numerous population studies of pulsars and their DM variations \citep[e.g.,][]{phillips1991, you2007, keith2013, jones2017, donner2020}. Most of these studies have focused on millisecond pulsars (MSPs), whose exceptional rotational stability makes them ideal for high-precision timing experiments. However, the LoSs to MSPs primarily sample the large-scale, diffuse IISM far from active star-forming regions. Young pulsars occupy a fundamentally different regime. They are often still embedded in or near their birth SNRs and surrounding pulsar wind nebulae (PWNe), where the local electron density and turbulence can exceed the ambient IISM values by orders of magnitude. Their DM variations therefore carry imprints of both the diffuse IISM and these dense, structured local environments, providing access to turbulence on spatial scales and in physical conditions that MSP studies cannot probe.

Despite this potential, the DM variations of young pulsars remain poorly characterized on a population level. \citet{petroff2013} analyzed 168 young pulsars observed at 0.7, 1.4, and 3.1~GHz with a 40, 1024, and 256~MHz bandwidth over five years at the Murriyang/Parkes radio telescope but, because of the significant uncertainties in their DM estimates, found significant DM variations in only four objects. More recently, \citet{keith2024} studied 597 pulsars over four years with the radio telescope MeerKAT and identified 87 pulsars with significant DM gradients. Such gradients can be a signature of a Kolmogorov-like energy cascade, in which the power-law nature of the electron density fluctuations produces correlated, quasi-linear DM trends. These results raise a natural question: to what extent are the DM variations of young pulsars driven by Kolmogorov turbulence in the diffuse IISM, and to what extent by the dense, structured environments in which some of these pulsars reside? Answering this question requires both a large sample of young pulsars and a long observational baseline with sufficient bandwidth to measure DM variations combined with the precision needed to distinguish between these scenarios.

In this paper, we present one of the largest and longest DM variation studies of young pulsars to date. We analyzed 364 young pulsars observed with the Murriyang/Parkes radio telescope over more than 18 years, combining data from the 21~cm Multibeam receiver \citep{staveley1996} with recent observations from the ultra-wideband low-frequency receiver (UWL; \citealt{hobbs2020}). The UWL provides continuous frequency coverage from 704 to 4032~MHz, and its 3.3~GHz instantaneous bandwidth reduces per-epoch DM uncertainties by an order of magnitude relative to the previous multibeam system. This combination of a long timing baseline with wideband precision allowed us to detect subtle and long-term DM variations that were inaccessible to earlier studies and to characterize the turbulence spectrum along individual LoSs.

This paper is structured as follows. Sect.~\ref{sec:methodology} describes the observations and analysis methods, including the DM extraction and structure function (SF) estimation. Sect.~\ref{sec:results} presents the population-level results and detailed case studies of individual pulsars with notable LoSs. We conclude in Sect.~\ref{sec:conclusions}.

\section{Observations and data analysis}\label{sec:methodology}
\begin{figure*}
    \centering
        \includegraphics[width=\hsize]{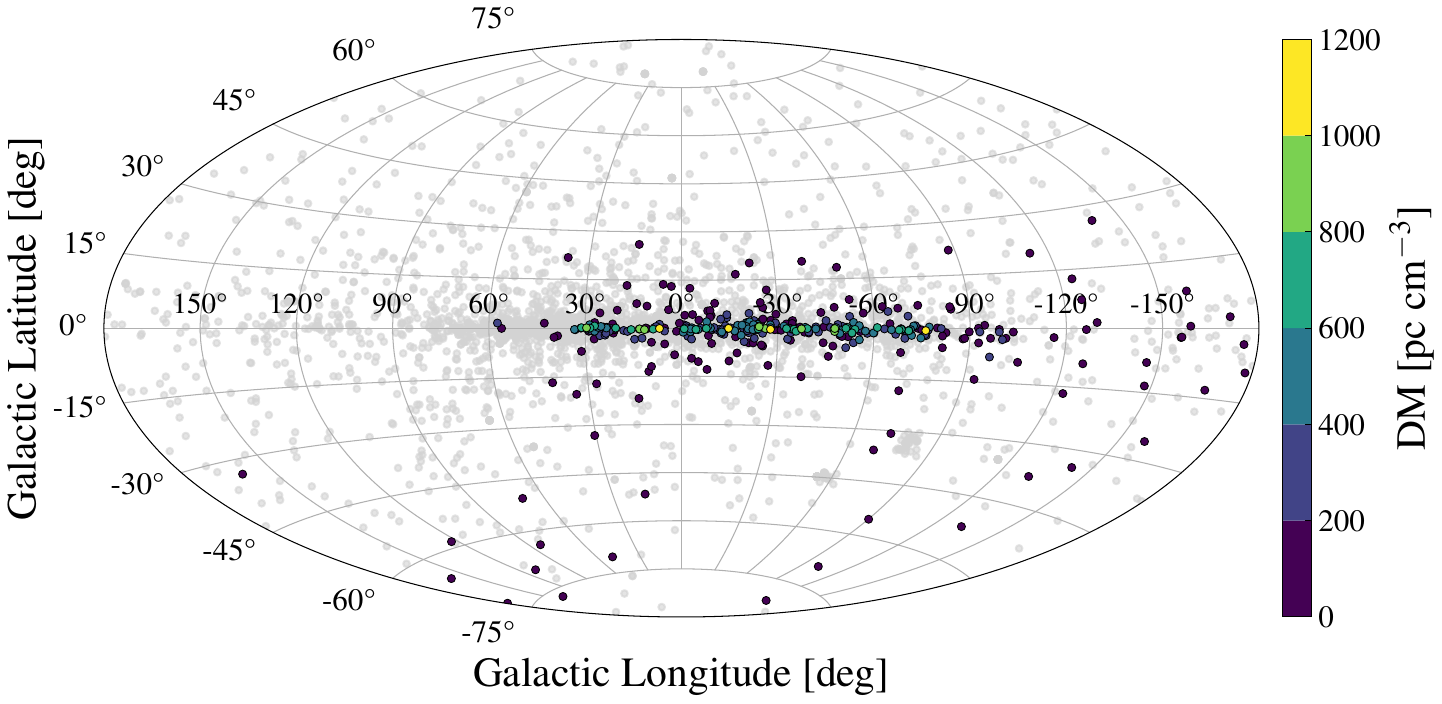}
    \caption{Aitoff sky projection map of known pulsars in Galactic coordinates. The background gray points represent the full sample of known radio pulsars in \textsc{psrcat}, while colored markers denote the subset of pulsars analyzed in this work. The marker color encodes the value of the DM, as shown by the color bar.}
\label{fig:dm_aitoff_map}
\end{figure*}

The pulsar observations analyzed in this work were collected by the Murriyang/Parkes radio telescope between 2007 and November 2025 as part of the young pulsar timing project \citep{weltevrede2010, johnston2021}. The sample contains 364 pulsars, with DM values ranging from $\sim 3$~pc~cm$^{-3}$ (PSR~J0108--1431) to $\sim 1160$~pc~cm$^{-3}$ (PSR~J1616--5109). The location of our pulsar sample on the sky is shown in Fig.~\ref{fig:dm_aitoff_map} alongside the full ATNF Pulsar Catalogue \citep{psrcat}.\footnote{\textsc{psrcat}, version 2.7.0, \url{https://www.atnf.csiro.au/research/pulsar/psrcat/}} Each pulsar was observed approximately once a month, with a typical integration time of $\sim 200$~s per epoch.

Two primary receiver systems were used over the course of the program. Before November 2018, observations were made with the 21~cm Multibeam Receiver at a center frequency of 1369~MHz and a bandwidth of 256~MHz, yielding 325 pulsars observed with this receiver. We excluded observations made at center frequencies of 0.7 and 3.1~GHz \citep{granet2005} with a six-month cadence, as their sparse temporal sampling is insufficient for the SF analysis described below. Observations made with the H-OH receiver \citep{granet2011} are also excluded for the same reason. From November 2018 onward, the UWL receiver was installed and after August 2021 solely used to observe our pulsar sample, which provides continuous frequency coverage from 704 to 4032~MHz at a resolution of 1~MHz. The UWL data cover 291 pulsars, of which 261 were also observed with the multibeam receiver; this enabled us to combine the two datasets for these pulsars into a single time series across the full observing time span.

\subsection{Template creation and timing}\label{sec:template_timing}
We analyzed the multibeam and UWL datasets separately but followed the same prescription to create the template of the pulse profile used to measure the times of arrival (ToAs). To create the template for each pulsar, we time-averaged a significant number of high signal-to-noise ratio (S/N) observations, which were chosen based on the median S/N from the archives, and used the \textsc{psradd} tool in \textsc{psrchive} \citep{hotan2004}. We ensured profile alignment across the entire dataset by using the predicted phase from \textsc{tempo2} \citep{hobbs2006} to rotate the archives before time-averaging. To determine the number of frequency channels to use, we weighted the channels used by the median S/N and the bandwidth of the archives. A similar routine\footnote{\url{https://github.com/nf-core/meerpipe/blob/dev/modules/local/dm_rm_calc.nf}} is implemented in the \textsc{meerpipe} package used by the MeerKAT Pulsar Timing Array \citep{miles2023}. We limited the maximum number of frequency channels that were used to 32 and 256 channels for the multibeam and UWL data, respectively. The frequency and time-averaging were performed using the \textsc{pam} routine in \textsc{psrchive}. Finally, the data-derived templates were then smoothed using the wavelet-smoothing technique in \textsc{psrsmooth}.

We correlated the template with the frequency-dependent, time-averaged observations using the Fourier-domain Markov-chain Monte Carlo method \citep{taylor1992,arzoumanian2015} in \textsc{pat} to estimate our ToAs. In this method, every frequency channel of the observation is matched with the corresponding channel of the template, which results in sub-banded ToAs for every observation. We then removed the outlier ToAs using a Huber-regression-based rejection scheme outlined in \citet{tiburzi2019}. 

\subsection{Obtaining DM time series and structure functions}\label{sec:dm_timeseries_structurefunctions}
After creating the frequency-resolved ToAs, we fit a DM value and a constant timing offset in our ToAs at each epoch using \textsc{tempo2}. Fitting per-epoch DM values directly to the sub-banded ToA using Eq.~\ref{eq:t_dm} avoids the need for a phase-connected timing solution or any modeling of the achromatic red noise in the timing residuals \citep{iraci2024}, both of which can be challenging for young, energetic pulsars with substantial spin noise. The result is a single DM estimate with its associated 1$\sigma$ uncertainty for every observation.

\subsubsection*{DM slope as a first-order diagnostic for DM variations}
As a first-order characterization of the DM variability in our sample, we performed a reduced chi-squared linear fit to each DM time series and quantified the significance of the trend using the fitted slope and its associated uncertainty. We use the absolute slope as a proxy for the amplitude of DM variations. Following this approach, we considered a DM slope to be significant if it exceeds the $5\sigma$ level. As noted by \citet{petroff2013}, a linear fit is insensitive to variations occurring on timescales shorter than the data span, but it provides a robust and uniform statistic for identifying pulsars with significant DM variations across the full sample.

\subsubsection*{DM structure functions}
A more complete characterization of the DM variability is provided by the DM SF, which measures the average squared difference in DM as a function of the time lag between measurements,
\begin{equation}
    D_{\rm DM}(\tau) = \left\langle
    \left[\text{DM}(t + \tau) - \text{DM}(t)\right]^2 \right\rangle,
    \label{eq:sf}
\end{equation}
where $\left< \cdots \right>$ denotes an ensemble average and $\tau$ is the time lag that separates two DM measurements. The SF encodes the turbulence spectrum of the IISM along the LoS: in the intermediate lag regime, where the data span is shorter than the outer scale of the turbulence, the SF follows a power law whose slope is directly related to the spectral index $\beta$ of the electron density wavenumber spectrum (see Eq.~\ref{eq:spectrum}). For a Kolmogorov medium with $\beta = 11/3$, the SF scales as $\tau^{5/3}$. At the smallest lags, the SF is dominated by the measurement noise floor, and at the largest lags, it saturates as the time lag approaches the outer scale of the turbulence. In full, the observed SF can be decomposed as \citep[e.g.,][]{lam2016,jones2017}
\begin{equation}
    D_{\rm DM}(\tau) = D^{\rm{sys}}_{\rm DM}(\tau) + D^{\rm{pl}}_{\rm DM}(\tau) + D^{\rm noise}_{\rm DM}
\end{equation}
where the three terms correspond to systematic trends, a turbulent power-law component, and a white-noise floor, respectively. The systematic term captures any linear and periodic structure in the DM time series. A dominant linear trend produces a $\tau^{2}$ ($\beta = 4$) dependence above the measurement noise floor \citep[e.g.,][]{lam2016}, while periodic modulations appear as dips at harmonics of the corresponding period, which can be caused by annual variations in the solar wind or the ionosphere \citep[e.g.,][]{tiburzi2019}. 

The SF can be extended to shorter time lags than those sampled by the DM time series by anchoring it to the diffractive interstellar scintillation (DISS) timescale, $\Delta\tau_{\rm d}$, measured at an observing frequency $\nu$. The value of the SF extrapolated from the DISS timescale at time lag $\tau$ is \citep[e.g.,][]{lam2016}
\begin{equation}
    D_{\rm DM}(\tau) = \left(\frac{K\nu}{2\pi}\right)^2 \left(\frac{\tau}{\Delta\tau_{\rm d}}\right)^{\beta - 2},
    \label{eq:sf_diss}
\end{equation}
where $K \equiv 2.410 \times 10^{-4}~\text{MHz}^{-2}~\text{cm}^{3}~\text{pc}~\text{s}^{-1}$ is the dispersion constant. If literature values of $\Delta\tau_{\rm d}$ are available for the individual pulsars discussed in Sect.~\ref{sec:psrs-individual}, we used Eq.~\ref{eq:sf_diss} to anchor the SF at sub-day lags; this provided a consistency check between the scintillation and DM variation regimes.

\subsubsection*{Structure function error estimation}
As the time lag approaches the length of the dataset, the limited number of independent realizations in the data limits the statistical reliability of the SF \citep{rickett2000}. Many young pulsars in our sample lack phase-connected timing solutions, making standard noise modeling with a Bayesian timing pipeline difficult. We therefore adopted a two-step approach to estimating the SF uncertainties.

First, we modeled the SF uncertainties that arise from the uncertainties on the estimated DM values. For each pulsar, we generated 10000 realizations of the DM time series by drawing each epoch independently from a Gaussian distribution centered on the estimated DM value with a standard deviation equal to its measurement uncertainty. We computed the SF for each realization at the sampled time lag and took our SF error bars to be the standard deviation of the resulting simulated SFs at each lag. 

Second, we modeled the SF uncertainties that arise from the assumption that the DM variations are described by a power-law process (see Eq.~\ref{eq:spectrum}). We first performed a least-squares fit to the observed SF, modeling it as the sum of a power-law component and a white-noise floor, $D_{\rm DM} = C\tau^{\beta - 2} + 2\sigma^2$ \citep[see Appendix~B of][]{lam2016}. We used this fit to calibrate the amplitude and spectral index for our simulated DM time series: for each pulsar, we generated 10000 realizations drawn from this power-law spectrum and took the 68\% confidence interval of the resulting SFs as the uncertainty at each time lag.

Finally, we refit the observed SF to the same model with the simulation-derived uncertainty band using the same least-squares fit. Additionally, we performed a Bayesian fit to the same model using the nested sampling algorithm \textsc{dynesty} \citep{speagle2020} as implemented in \textsc{bilby} \citep{ashton2019} and sampled the parameters with a log-uniform and uniform prior $\log_{10}C \in [-12, 4]$ and $\beta \in [0,7]$. We note that both these prior ranges are deliberately chosen to be wide to avoid biasing the fit and should be regarded as conservative bounds rather than physically motivated constraints. As discussed in Sect.~\ref{sec:introduction}, the observationally expected range for $\beta$ is much narrower than the prior range adopted here. Similarly, the broad prior range for $\log_{10}C$ far exceeds the range of values for a power-law amplitude expected for our data. For $\log_{10}\sigma$, we used a truncated Gaussian prior centered around the value inferred from the direct fit with a standard deviation of 2 and truncated to the same range as $\log_{10}C$. This prior is chosen because the constraints on $\log_{10}\sigma$ significantly impact those on $\beta$. The least-squares and Bayesian fits yield consistent results in the scintillation regime when $\beta$ is constrained in the Bayesian fit. We find that the uncertainty estimates on the parameters from the least-squares fit, derived from the diagonal of the covariance matrix alone, are significantly lower than those derived from the parameter posteriors of the Bayesian fit. We attribute this to the fact that the least-squares method does not capture the full degeneracy between the parameters, for example, between the amplitude and the spectral index, and the power-law amplitude and the white noise floor at the shortest lags. The Bayesian method samples the full parameter space and therefore captures these degeneracies, yielding more conservative uncertainties. Throughout the main text, we quote the uncertainties from the Bayesian fit. For completeness, both sets of uncertainties are listed in Appendix~\ref{app:significant_pulsars}.

We note that these simulations only model a stochastic power-law process and do not include a deterministic linear DM gradient. Although a power-law spectrum stochastically induces apparent linear trends over the finite span of our dataset, a gradient can arise from the changing distance to the pulsar due to the motion parallel to the LoS. Such a gradient contributes a $\tau^2$ term to the SF, both raising the overall SF values and steepening the measured slope \citep[e.g.,][]{lam2016}. Therefore, the inferred spectral index $\beta$ should be interpreted with caution, as it may include contributions from the motion of the pulsar rather than the underlying turbulent spectrum alone.

\section{Results and discussion}\label{sec:results}
We detect significant DM variations at the $\geq 5\sigma$ level in 89 pulsars in the UWL data, 86 of which are direct detections based on the significance of the DM gradient measured from the per-epoch DM estimates; the remaining 3 required additional modeling (Appendix~\ref{app:dm_jitter_pulsars}), as well as in 23 pulsars in the multibeam data. There are 16 pulsars that also show a significant DM slope detection in the combined dataset. The full results are tabulated in Appendix~\ref{app:significant_pulsars}. This represents a substantial improvement over the four detections reported by \citet{petroff2013} for the same telescope, which we attribute to two main factors: our data time span exceeds theirs by more than a factor of two, and the 3.3~GHz instantaneous bandwidth of the UWL reduces per-epoch DM uncertainties by approximately an order of magnitude relative to the 256~MHz multibeam system. Our measurements are consistent with theirs over the overlapping data span.

\subsection{Excluded pulsars}\label{sec:excluded}
Not all pulsars with significant DM gradients enter the count above. We excluded two pulsars whose apparent DM variations are driven by profile shape changes rather than IISM propagation effects. PSR~J1428--5530 (B1424--55) shows pulse-width variations most prominent in the trailing edge of the profile \citep{lower2025}, while PSR~J1825--0935 (B1822--09) exhibits correlated changes in the interpulse and main pulse amplitude with the same period as the spin frequency \citep{yan2019}. PSR~J1105--6107 and PSR~J1829--1751 show periodic profile evolution but are included in the sample because the linear component in their DM variations can be disentangled from the periodic variations since the periodic variations are approximately symmetric around the linear trend and therefore largely average out over time (see Sect.~\ref{sec:J1105-6107_J1829-1751} for a full discussion). More broadly, intrinsic profile evolution across the UWL bandwidth is a potential source of bias in DM estimates as the pulse profile can vary significantly across the full bandwidth. To mitigate this effect, we used frequency-resolved templates of each pulsar (see Sect.~\ref{sec:template_timing}), such that the template at each sub-band captures the profile shape at that observing frequency.

\begin{figure}
    \centering
    \includegraphics[width=\hsize]{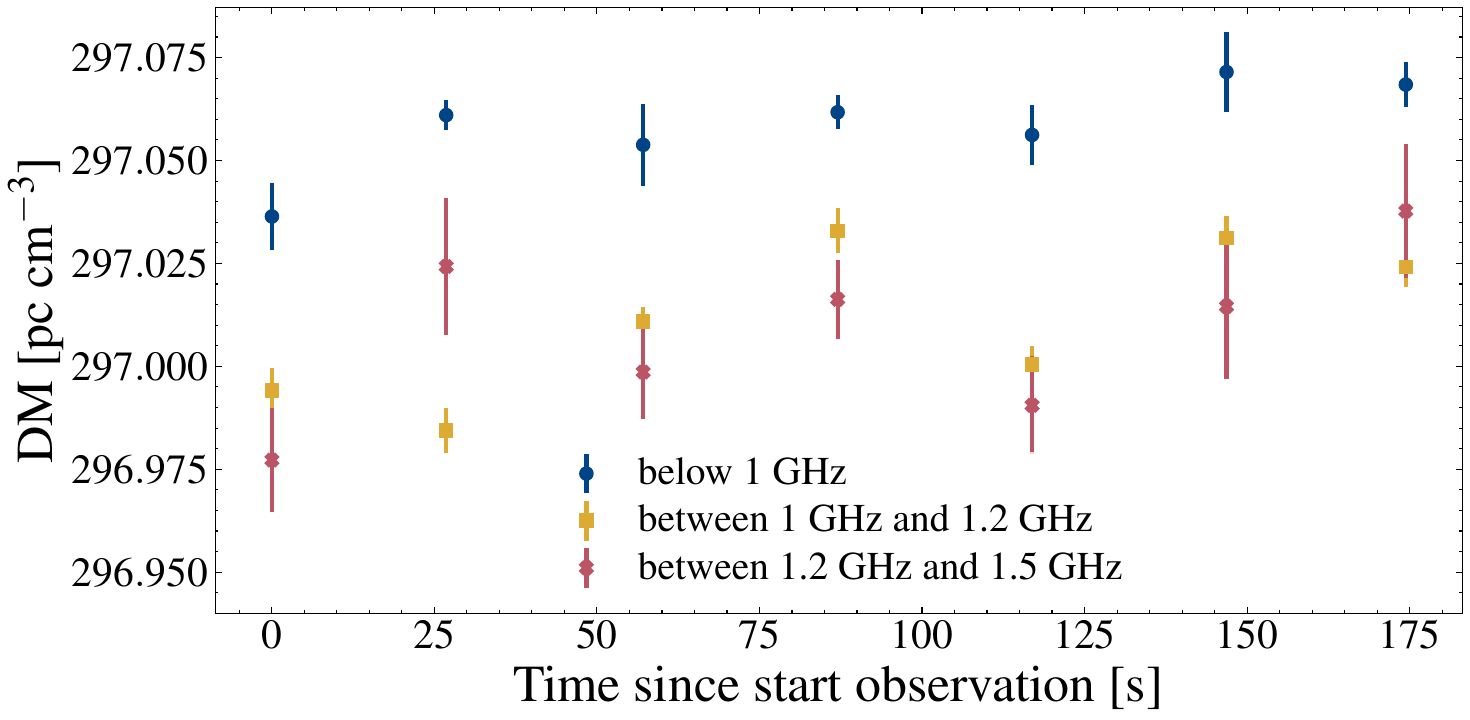}
    \caption{DM time series for the lowest three (out of five) frequency sub-bands of one epoch (MJD 60414) for PSR~J1243--6423. Each point corresponds to a 30~s sub-integration. The blue circles, yellow squares, and pink crosses correspond to the below 1~GHz, 1--1.2~GHz, and 1.2--1.5~GHz bands, respectively. Error bars indicate the measurement uncertainties for each DM estimate.}
    \label{fig:subband_dm}
\end{figure}

A further 21 pulsars show $\geq 3\sigma$ DM slopes in the UWL data but are also excluded from the main sample because their epoch-to-epoch DM scatter significantly exceeds the level expected from IISM turbulence. We identified these pulsars based on two criteria: the epoch-to-epoch variations greatly exceed the DM uncertainties, and the affected pulsars are all bright ($\text{S/N} \gtrsim 500$) with spin periods $P \gtrsim 0.3$~s. Both properties point to pulse jitter as the dominant noise source rather than genuine DM variability driven by IISM turbulence. To still obtain a first-order estimate of the DM variations, we inflated the DM uncertainties with an additional white-noise term in quadrature (EQUAD). Using these uncertainties, we find three additional pulsars with a significant DM slope (see Appendix~\ref{app:dm_jitter_pulsars} for details). This brings the total number of pulsars with significant DM variations in the UWL data to 89 pulsars.

The excess scatter is explained by pulse jitter, which arises from stochastic pulse-to-pulse variations in the profile shape and amplitude. Jitter is a broadband phenomenon that does not average down with increasing bandwidth \citep{cordes1985}. It does decrease with the number of pulses, typically scaling with the number of pulses $N_{\rm p}$ as $\sigma_{\rm J} \propto N_{\rm p}^{-1/2}$, so shorter integrations and longer spin periods are most affected \citep{shannon2014}. With our observing program's $\sim 200$~s integrations, a pulsar with $P = 0.5$~s accumulates only $\sim 400$ pulses per epoch, too few to suppress jitter to the level of IISM-induced DM variations. PSR~J1243--6423 illustrates this: Fig~\ref{fig:subband_dm} shows the DM values per individual sub-band for a single observation from each 30~s sub-integration. On these short timescales, we see that the DM time series per sub-band show nonidentical structure. This behavior is not unique to this observation: for a set of nine archives, the average correlation between the three sub-bands shows that adjacent frequency bands show moderate correlation ($\sim 0.4$), while the lowest and highest frequency bands are uncorrelated ($\sim 0.08$).  

Pulse jitter has been shown to de-correlate in bandwidths much smaller than the 3.3~GHz bandwidth of the UWL \citep[e.g.,][]{shannon2012,lam2019, parthasarathy2021}. As such, the observed de-correlation shown for PSR~J1243--6423 can be explained by intrinsic pulse-to-pulse variability. This effect presents a fundamental limit on the achievable DM precision and therefore on large-scale studies of the IISM.

The UWL exacerbates this effect in a way that the multibeam system does not. The order-of-magnitude reduction in DM uncertainty brought about by the wider bandwidth and lower minimum observing frequency reveals DM misestimation from jitter that was previously hidden below the radiometer noise floor. Longer integrations would mitigate the observed short-term variations but are incompatible with the project design, which distributes telescope time across hundreds of pulsars. This represents a fundamental trade-off between sample size and per-pulsar timing quality that will become increasingly relevant in the Square Kilometre Array (SKA) era (see Sect.~\ref{sec:conclusions}).

\subsection{The |dDM/dt| versus DM relation}\label{sec:dm_slope}
\begin{figure}
    \centering
    \includegraphics[width=\hsize]{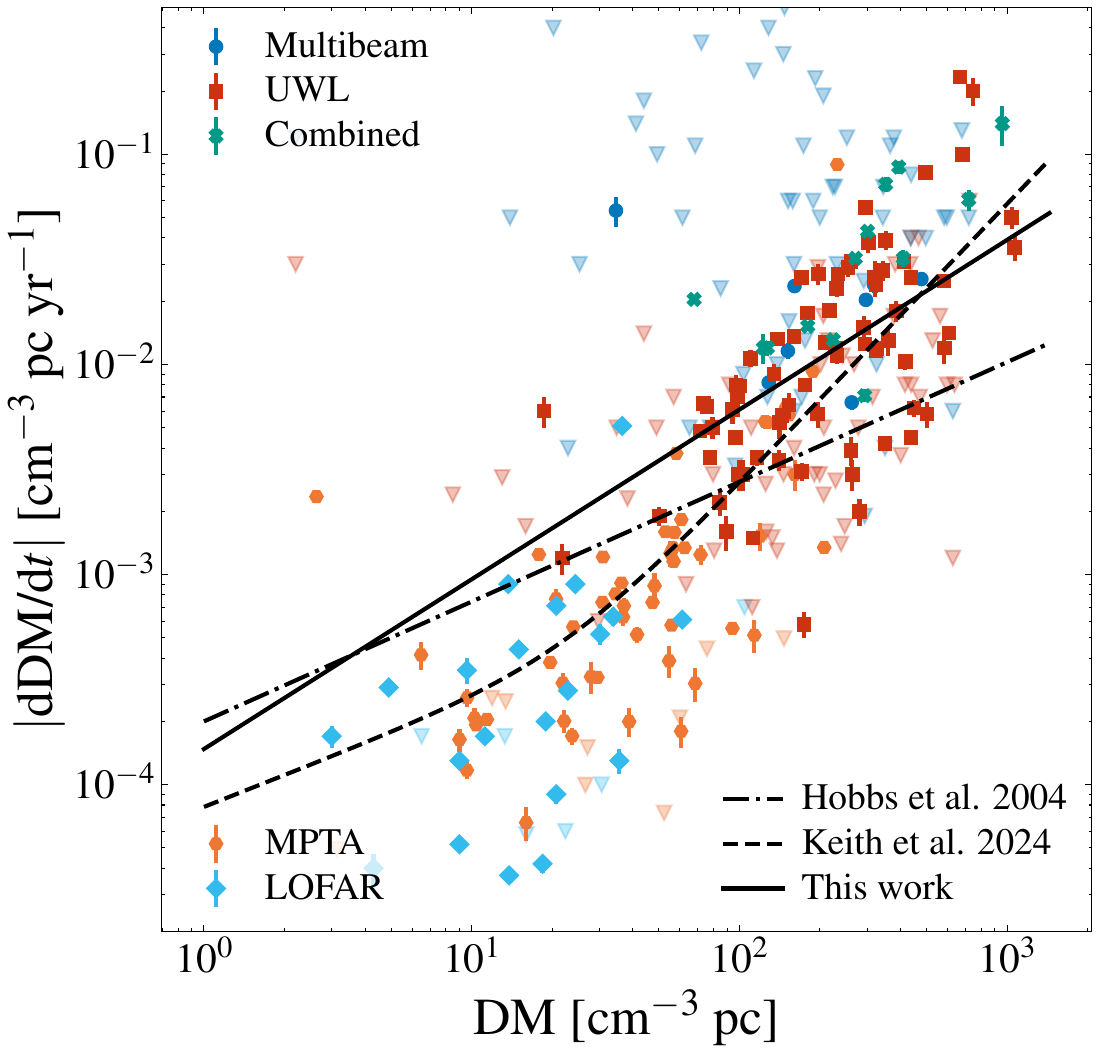}
    \caption{Absolute DM gradient as a function of DM. Points show constraints from the UWL (red), multibeam (blue), and combined (green) datasets. MSPs observed with LOFAR \citep{donner2020} and MeerKAT \citep{miles2023} are shown in orange and cyan, respectively. The downward triangles indicate $2\sigma$ upper limits. The dash-dotted line shows the power-law relation derived by \citet{hobbs2004}, the dashed line the relation derived by \citet{keith2024}, and the solid line the relation in Eq.~\ref{eq:pl_fit}. Upper limits lying well above the plotted range are excluded for clarity.}
    \label{fig:dm_vs_slope}
\end{figure}

\citet{hobbs2004} found, from 374 pulsars spanning up to 32 years of timing data, that the DM slope scales approximately as $|\text{dDM}/\text{d}t| \propto \text{DM}^{1/2}$. In their sample, 21 pulsars show a DM slope that is above the $3\sigma$ level. Their slope is consistent with a LoS that samples a turbulent IISM whose DM variation amplitude grows with path length \citep{backer1993}. Fitting a power law of the same form to our $\geq 5\sigma$ detections, we find that
\begin{equation}
    \left|\frac{\text{dDM}}{\text{d}t}\right| = (1.5\pm 1.0) \times 10^{-4} \times \text{DM}^{0.81\, \pm \,0.13} \text{ cm}^{-3}\text{ pc yr}^{-1},
    \label{eq:pl_fit}
\end{equation}
which is a steeper slope than the \citet{hobbs2004} relation, as they find a slope of $0.57\pm 0.09$.

\citet{donner2020} also found a steeper slope than \citet{hobbs2004} from 7 years of the Low-Frequency Array (LOFAR) observations of 32 MSPs, attributing the discrepancy to large uncertainties in the \citet{hobbs2004} sample and the inclusion of Galactic-center pulsars with extreme DMs. \citet{keith2024} derived a relation between the DM slope and DM using IISM scattering measurements from \citet{krishnakumar2015}. Our relation is more consistent with \citet{keith2024} than with \citet{hobbs2004}, as shown in Fig.~\ref{fig:dm_vs_slope}, although the scatter around the fitted line is substantial and the simple power-law model is not easily distinguished by our data alone. We note that the slope in our fit is somewhat flattened by a few low-DM pulsars that have high DM gradients. If we restrict our fit to pulsars that have $\text{DM} > 100~\text{cm}^{-3}~\text{pc}$, we find a steeper slope of $1.07 \pm 0.20$. This is in closer agreement with the \citet{keith2024} model at high DM values. Since the LOFAR and MeerKAT data are excluded from our fit, our power-law model does not leverage those low-DM points. For completeness, Fig.~\ref{fig:dm_vs_slope} also includes DM gradients from the first MeerKAT Pulsar Timing Array data release \citep{miles2023}.

The majority of our pulsars with detected DM variations are consistent with a Kolmogorov wavenumber spectrum ($\beta = 11/3$) within $5\sigma$ (see Appendix~\ref{app:significant_pulsars}): out of 38 pulsars in the UWL dataset with a constrained $\beta$ in the Bayesian fit, 31 are consistent with a Kolmogorov spectral index. For the multibeam and combined dataset, we find that this is 7 out of 8 and 5 out of 6 pulsars. In the least-squares fit, we find that this is the case for 4, 22, and 4 pulsars for the multibeam, UWL, and combined datasets, respectively. However, due to the uncertainties on $\beta$, a substantial fraction of these are also consistent with $\beta = 4$ in both methods, the value expected when the DM SF is dominated by a linear DM gradient: this is the case for 28 of the 38 UWL pulsars, 3 of the 8 multibeam pulsars, and 5 of the 6 combined dataset pulsars. We note that we did not model the contribution to the SF that arises from the linear gradient itself, which introduces a bias in $\beta$ inferred from the SF slope (see Sect.~\ref{sec:dm_timeseries_structurefunctions}). The individual pulsars with the most notable LoS are discussed in Sect.~\ref{sec:psrs-individual}.

\subsection{Scattering as a source of DM bias}\label{sec:scattering}
A second chromatic effect present in our data is pulse broadening due to multipath scattering in the IISM. A scattered pulse often exhibits an exponential tail with a characteristic timescale $\tau_{\rm s} \propto \nu^{-\alpha}$, where $\alpha = 4$ for a thin scattering screen \citep{cronyn1970,lang1971} and $\alpha = 4.4$ for a Kolmogorov medium \citep{lee1976, rickett1977}. Because the scattering delay itself is frequency-dependent, it can bias sub-banded ToAs and thus contaminate per-epoch DM measurements \citep[e.g.,][]{geiger2025}.

We modeled the pulse broadening function as a convolution of the intrinsic (assumed Gaussian) profile with an isotropic scattering function,
\begin{equation}
    p_{\rm d}(t) = \frac{e^{-t/\tau_{\rm s}}}{\tau_{\rm s}}\,\Theta(t),
\end{equation}
where $\Theta(t)$ is the Heaviside step function, and fit each frequency channel independently using a Markov chain Monte Carlo algorithm to obtain scattering timescale per frequency channel. To assess the magnitude of this bias in our data, we measured $\tau_{\rm s}$ and $\alpha$ per epoch for PSR~J1512--5759 with a clearly visible scattering tail, following the method of \citet{geyer2017} and \citet{oswald2021}; see Sect.~\ref{sec:J1512-5759} for the detailed analysis.

Refractive effects provide an additional potential source of chromatic bias. Large-scale density gradients in the IISM refract the radio beam, introducing angle-of-arrival variations and a geometric delay that does not follow the $\nu^{-2}$ dispersion law \citep[e.g.,][]{foster1990,cordes2010}. Because this geometric delay scales steeply with wavelength, it is strongly chromatic and can leak into per-epoch DM estimates. For a pulsar with a fiducial bending angle $\theta_r \sim 1~\mathrm{mas}$ and an effective distance $D_{\rm eff} \sim 3~\mathrm{kpc}$, the implied geometric delay is $\Delta t_{\rm geo} \sim 3.6~\mu\mathrm{s}$. Via Eq.~\ref{eq:t_dm}, this translates to a DM error of $\sigma_{\rm DM}\sim 10^{-4}~\mathrm{pc}~\mathrm{cm}^{-3}$ across the UWL. This is well below most of our pulsar's per-epoch DM uncertainties, so we expect this bias to be subdominant. However, for strongly scattered LoS such as in the direction PSR~J1512--5759 (see Sect.~\ref{sec:J1512-5759}), the refractive delay can approach the per-epoch DM uncertainty and may contribute at a low level to the epoch-to-epoch chromatic scatter, and we cannot fully exclude them as a contaminant alongside the scattering-delay leakage discussed above.

\subsection{Individual pulsars with notable DM variations}\label{sec:psrs-individual}

\subsubsection{PSR J0835--4510 and PSR J0908--4913}\label{sec:J0835-4510_J0908-4913}
PSR~J0835--4510 and PSR~J0908--4913 offer a rare opportunity to test whether IISM turbulence driven by SNR environments is consistent with a Kolmogorov energy cascade. Both pulsars are embedded in, or closely associated with, local SNRs and PWNe, placing their LoSs in environments where the electron density and turbulence amplitude are expected to far exceed the diffuse IISM.

PSR~J0835--4510 \citep[B0833--45;][]{large1968} has a spin period of 89~ms and a spin-down energy loss rate of $6.9 \times 10^{36}$~ergs~s$^{-1}$. The pulsar is located within the Vela SNR at a well-constrained distance of 290~pc \citep{caraveo2001} and drives the PWN Vela--X \citep{weiler1980}. The Vela SNR clump contributes significantly to the electron column density along this LoS.

PSR~J0908--4913 \citep[B0906--49;][]{damico1988} has a spin period of 106~ms and a spin-down energy loss rate of $4.9 \times 10^{35}$~ergs~s$^{-1}$. Its distance is less certain: H{\sc i} absorption measurements place an upper limit of 6.7~kpc \citep{koribalski1995}, while DM model-based distance estimates from YMW16 and NE2025 \citep{ocker2026} yield a lower limit of 1.0~kpc. We adopted 3~kpc, which is consistent with the pulsar's pulsed gamma-ray emission \citep{abdo2013} and with the revised environment analysis of \citet{johnston2021_2}. \citet{gaensler1998} detected a clear PWN and bow shock around PSR~J0908--4913; the implied average hydrogen density along the LoS is $n_{\rm H} > 2$~cm$^{-3}$ at 6.7~kpc, compared to $n_{\rm H} \approx 0.08$~cm$^{-3}$ at 3~kpc, the latter being consistent with typical IISM values \citep{johnston2021_2}. \citet{johnston2021_2} also found a possible association between PSR~J0908--4913 and the SNR candidate G320.4--1.0; if confirmed, this would place the pulsar's true age at $\sim$13~kyr, far below its characteristic age of 112~kyr.

\begin{figure*}
    \centering
    \begin{subfigure}[t]{0.49\textwidth}
        \includegraphics[width=\textwidth]{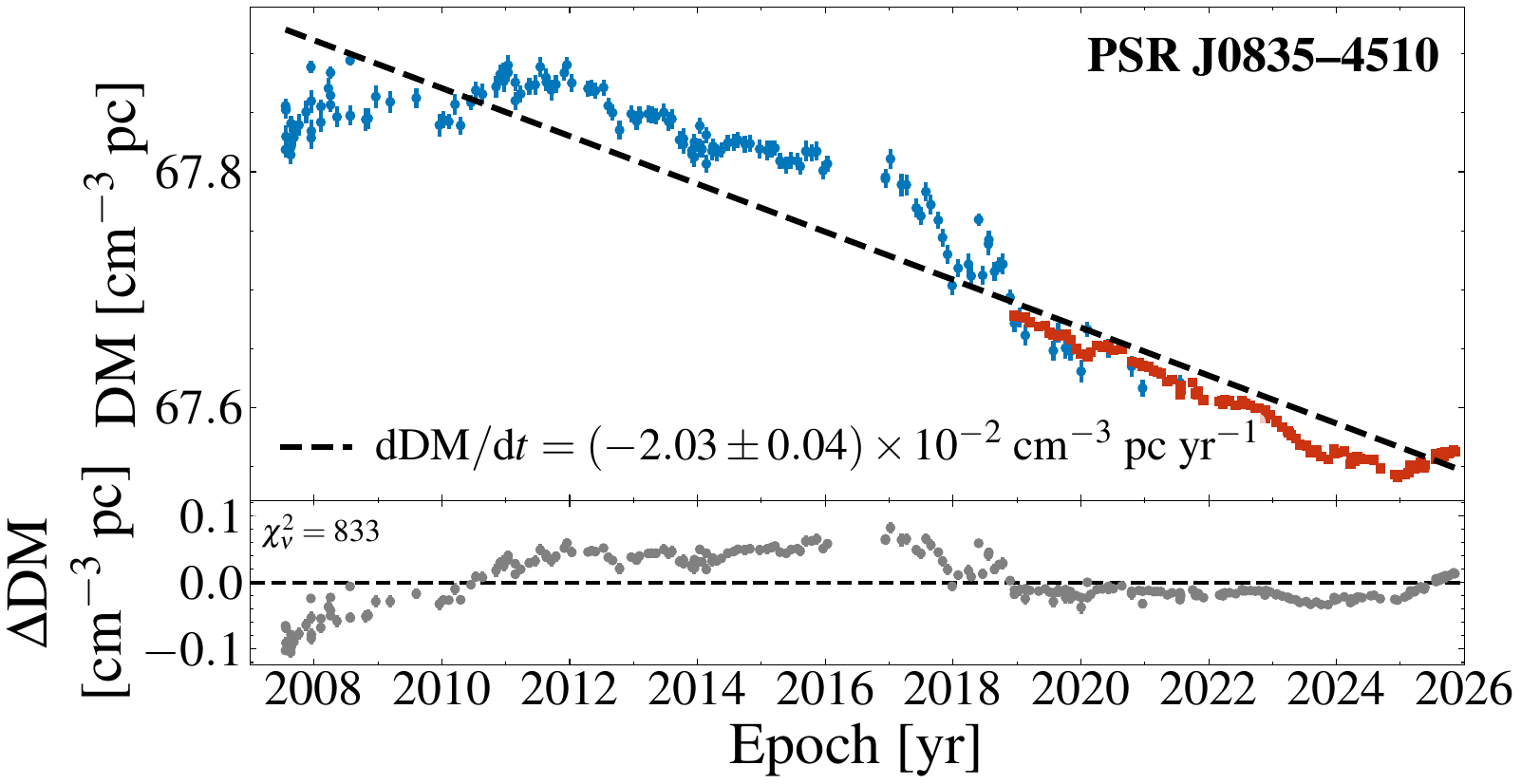}
    \end{subfigure}
    \hfill
    \begin{subfigure}[t]{0.49\textwidth}
        \includegraphics[width=\textwidth]{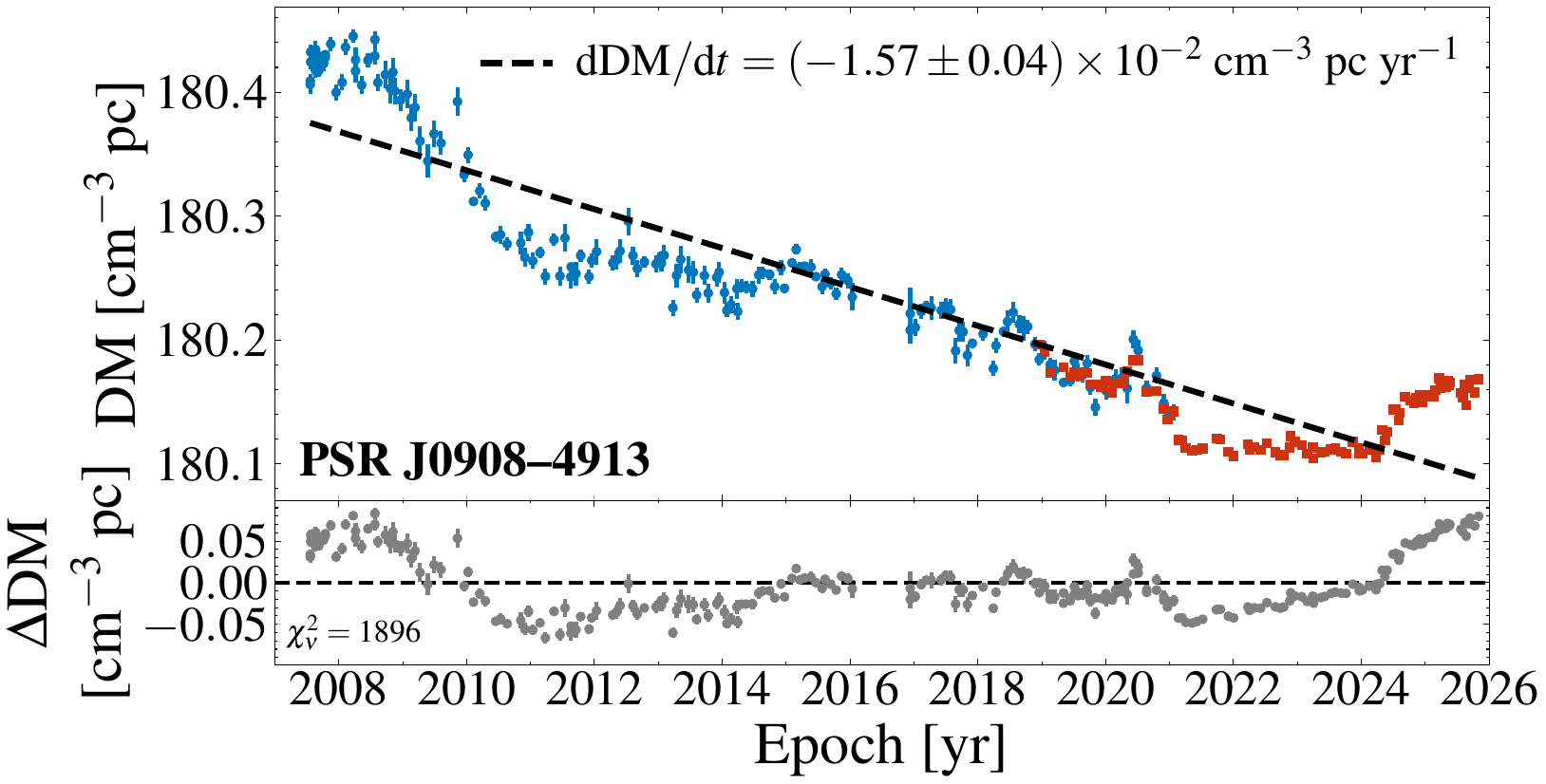}
    \end{subfigure}
    \vfill
    \vspace{1em}
    \begin{subfigure}[t]{0.49\textwidth}
        \includegraphics[width=\textwidth]{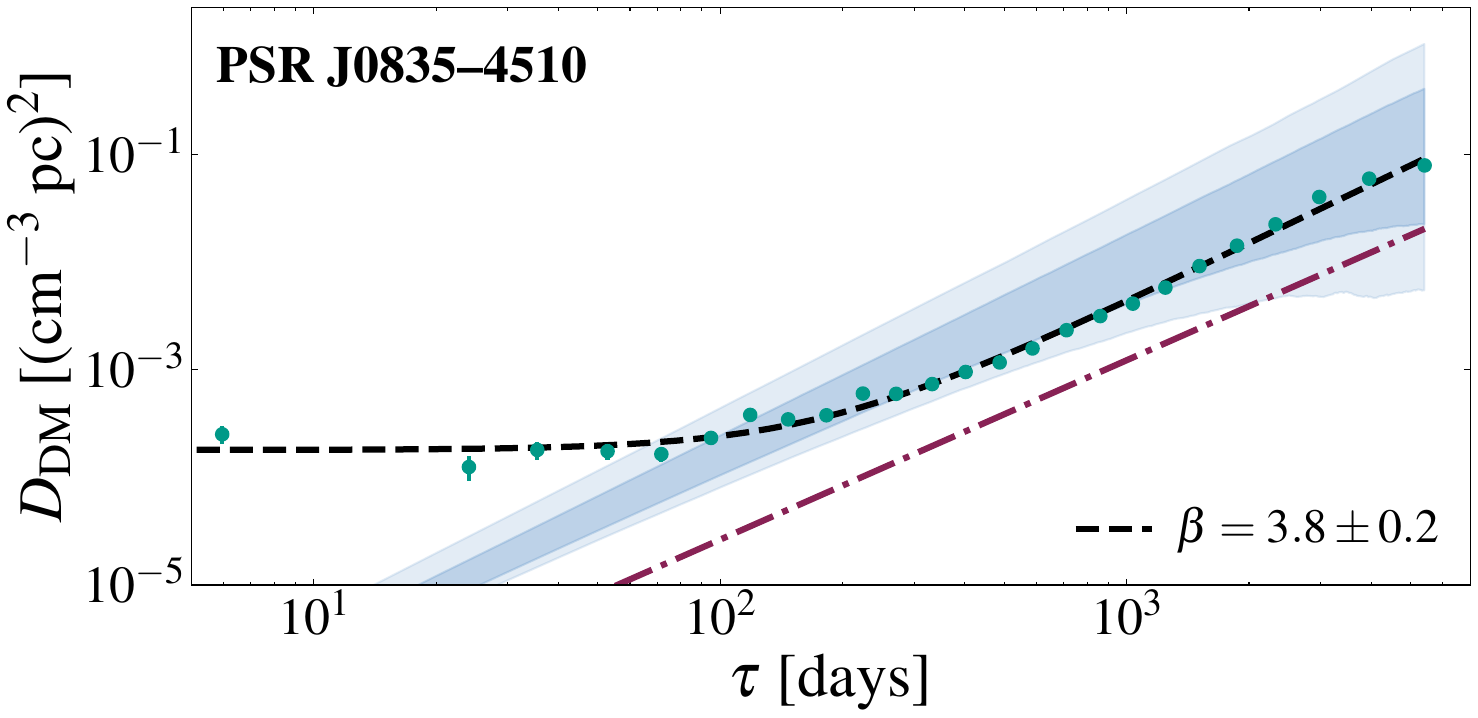}
    \end{subfigure}
    \hfill
    \begin{subfigure}[t]{0.49\textwidth}
        \includegraphics[width=\textwidth]{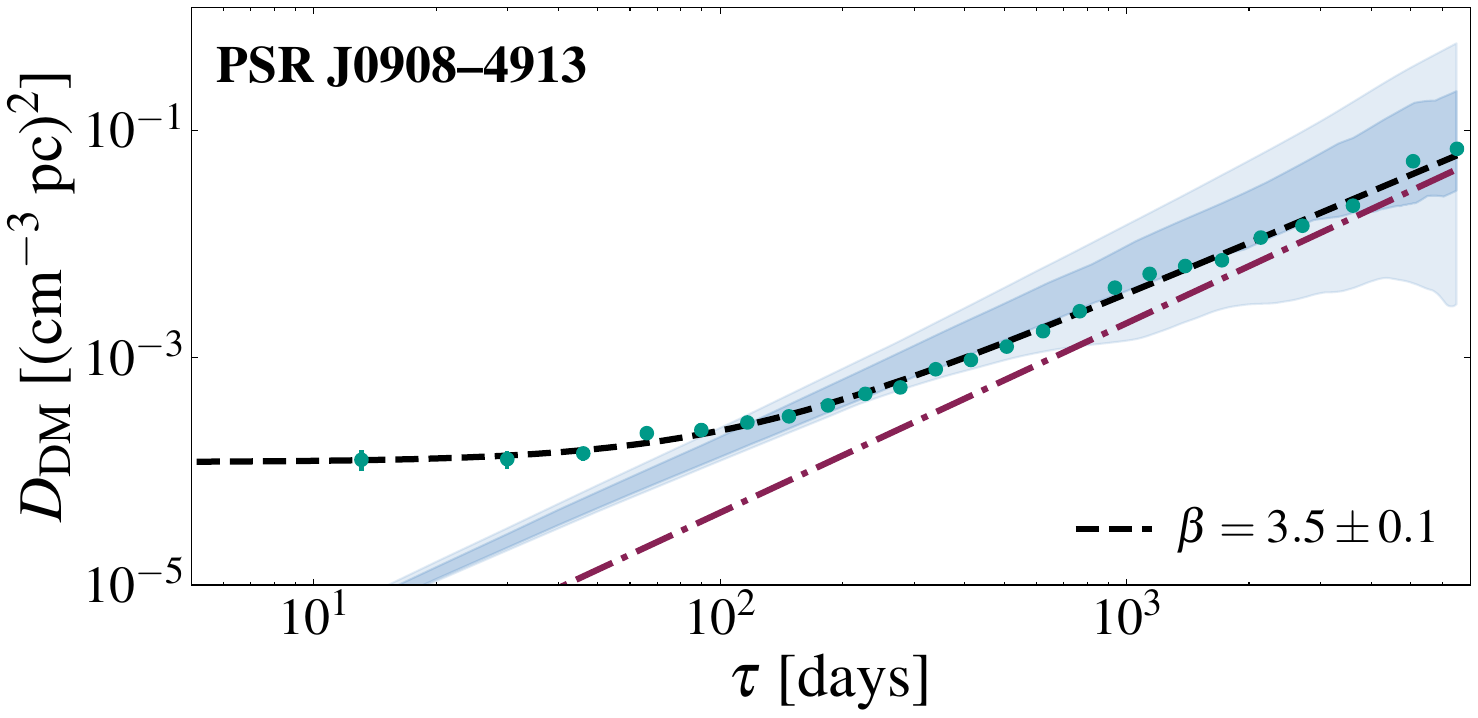}
    \end{subfigure}
    \caption{DM time series (\textit{top panels}) and DM SFs (\textit{bottom panels}) for PSRs~J0835--4510 (\textit{left}) and J0908--4913 (\textit{right}). The points in the top panels show the measured DM values with their corresponding $1\sigma$ uncertainties for the multibeam (blue) and UWL (red) datasets. The gray points in the bottom panels show the DM values after subtracting the fitted linear trend. The reduced chi-squared ($\chi_\nu^2$) is computed after removing the linear trend. The cyan points show the DM SF computed from the DM time series. The light blue shaded regions indicate the 68\% and 95\% confidence intervals derived from simulations of a power-law medium. The dashed black line shows the power-law fit with a constant noise floor based on the median values inferred from the Bayesian fit. The dashed-dotted dark red line shows the SF inferred from extrapolation of the DISS timescale assuming a Kolmogorov medium.}
\label{fig:J0835-4510_J0908-4913}
\end{figure*}

\paragraph{DM time series:}
The DM time series for both pulsars, spanning 18 years of combined multibeam and UWL data, are shown in the top panels of Fig.~\ref{fig:J0835-4510_J0908-4913}. Both pulsars show significant, time-correlated DM variations across both receiver systems.

For PSR~J0835--4510, we measure a decreasing DM slope of $(-2.03 \pm 0.04) \times 10^{-2}$~cm$^{-3}$~pc~yr$^{-1}$ over the full 18-year baseline. The overlapping multibeam data from \citet{petroff2013}, covering February~2007 to October~2012, exhibit a nearly flat DM trend, consistent with the shallow slope they reported. Looking further back, \citet{hamilton1985} found a much steeper decrease of $-0.040$~cm$^{-3}$~pc~yr$^{-1}$ from observations taken between 1970 and 1985, which they attributed to a magnetized filament moving out of the LoS within the Vela SNR. The flat structure seen in the \citet{petroff2013} data is consistent with this filament having moved fully out of the LoS by the mid-2000s. The varying slopes across these three epochs illustrate how the observed DM gradient for Vela reflects transient local structures rather than a steady bulk flow, underlining the importance of long baselines to correctly characterize the turbulence.

For PSR~J0908--4913, we find an overall DM slope of $(-1.57 \pm 0.04) \times 10^{-2}$~cm$^{-3}$~pc~yr$^{-1}$ from the combined dataset. The UWL data alone show a slope an order of magnitude smaller in absolute value because the UWL time series captures a decrease that flattens and reverses near the end of our data span. \citet{petroff2013} measured a decreasing slope of $-0.030 \pm 0.001$~cm$^{-3}$~pc~yr$^{-1}$, consistent with \citet{keith2024}. Rather than the dense PWN environment invoked by \citet{petroff2013} to explain the elevated DM variations, a simpler physical explanation is the association between PSR~J0908--4913 and SNR candidate G320.4--1.0: material from the SNR shell crossing our LoS would produce the observed filamentary DM variations.

\paragraph{Structure functions:}
To characterize the turbulence spectrum along both LoSs, we computed the DM SFs for the combined datasets, which are shown in the bottom panels of Fig.~\ref{fig:J0835-4510_J0908-4913}. At time lags shorter than those sampled by the DM time series, we anchored the SF using measured DISS timescales: $\Delta\tau_{\rm d} \sim 70$~s at 8.4~GHz for PSR~J0835--4510 \citep{johnston1998}, and $\Delta\tau_{\rm d} \sim 59$~s at 8.5~GHz for PSR~J0908--4913 \citep{johnston2021_2}.\footnote{ \citet{johnston1998} initially reported an incorrect scintillation timescale of 4770~s at 15.2~GHz for PSR~J0908--4913, which was later corrected in \citet{johnston2021_2}.} Fitting the SF with a power law and a constant noise floor, we find spectral indices of $\beta = 3.8 \pm 0.2$ and $\beta = 3.5 \pm 0.1$ for PSR~J0835--4510 and PSR~J0908--4913, respectively. Both values are consistent with the Kolmogorov expectation of $\beta = 11/3 \approx 3.67$ within $5\sigma$, although the median $\beta$ inferred for J0908--4913 is significantly shallower than expected from a Kolmogorov turbulence. 

The measured SF and our Kolmogorov simulation bands lie systematically above the extrapolation from the DISS timescale at all lags, rather than coinciding with it. Because our simulations model only the stochastic density-fluctuation component and do not include a deterministic linear trend (see Sect.~\ref{sec:methodology}), this offset can be explained by the fact that we did not model the SF contribution from the changing Earth-pulsar distance. We lack measurements of the parallel motion for J0835--4510 and J0908--4913, so we cannot directly quantify the distance-change contribution and thereby distinguish a quasi-linear trend of the wavenumber spectrum from contamination by a systematic trend. We further note that the scintillation timescale is itself expected to vary epoch-to-epoch as the LoS samples different IISM electron densities, so the DISS anchor point should be regarded with caution.

\paragraph{Turbulence amplitudes:}
Although the spectral slope is consistent with Kolmogorov turbulence, this does not exclude the possibility of a local source of enhanced fluctuations. To quantify this, we computed the wavenumber spectrum coefficient ($C_n^2$) from the SF,
\begin{equation}
    C_n^2 = \frac{1}{K_{\rm DM}}\left[\frac{D_{\rm DM}(\tau)}{D\,(v_{\rm eff,\perp}\,\tau)^{\beta-2}} \right],
    \label{eq:cn2_approx}
\end{equation}
where $K_{\rm DM} \simeq 88.3$ for $\beta = 11/3$, $D_{\rm DM}(\tau)$ is the SF value at time lag $\tau$, and $D$ is the Earth-pulsar distance \citep{cordes1998, lam2016}. Assuming that the dominant interstellar scattering screen lies near the pulsar (consistent with an SNR origin), we approximated $v_{\rm eff,\perp}$ with the pulsar's transverse velocity. We find $C_n^2 \simeq 10^{0.4 \pm 0.1}$~m$^{-20/3}$ for PSR~J0835--4510 and $C_n^2 \simeq 10^{-2.4 \pm 0.2}$~m$^{-20/3}$ for PSR~J0908--4913. Both values exceed the canonical level of $C_n^2 = 10^{-3.5}$~m$^{-20/3}$, which is the characteristic of typical LoSs through the local diffuse IISM \citep{cordes1998} and thus confirms enhanced turbulence along both LoSs \citep[e.g.,][]{johnston1998,chepurnov2010}. Our values are also comparable to the enhanced amplitude measured near bow shocks by \citet{ocker2021}, who found $C_n^2 \simeq 10^{-0.8}$~m$^{-20/3}$ for the Guitar Nebula and $C_n^2 \simeq 10^{-1.57}$~m$^{-20/3}$ for the very local ISM probed by Voyager~1. In both their case and ours, the elevated $C_n^2$ arises because the LoS passes through a region with enhanced density fluctuations compared to diffuse IISM.

The combination of a Kolmogorov spectral shape with an elevated $C_n^2$ supports a consistent physical picture: both pulsars reside in or near SNRs that inject turbulent energy at large scales, which then cascades to smaller scales and relaxes toward a Kolmogorov spectrum in the surrounding IISM. The DM variations along both LoSs are therefore most naturally explained by material within the SNR and PWN moving across the LoS, rather than by turbulence in the diffuse IISM far from the pulsar.

\subsubsection{PSR J1028--5819}\label{sec:J1028-5819}

\begin{figure}
    \centering
    \includegraphics[width=\hsize]{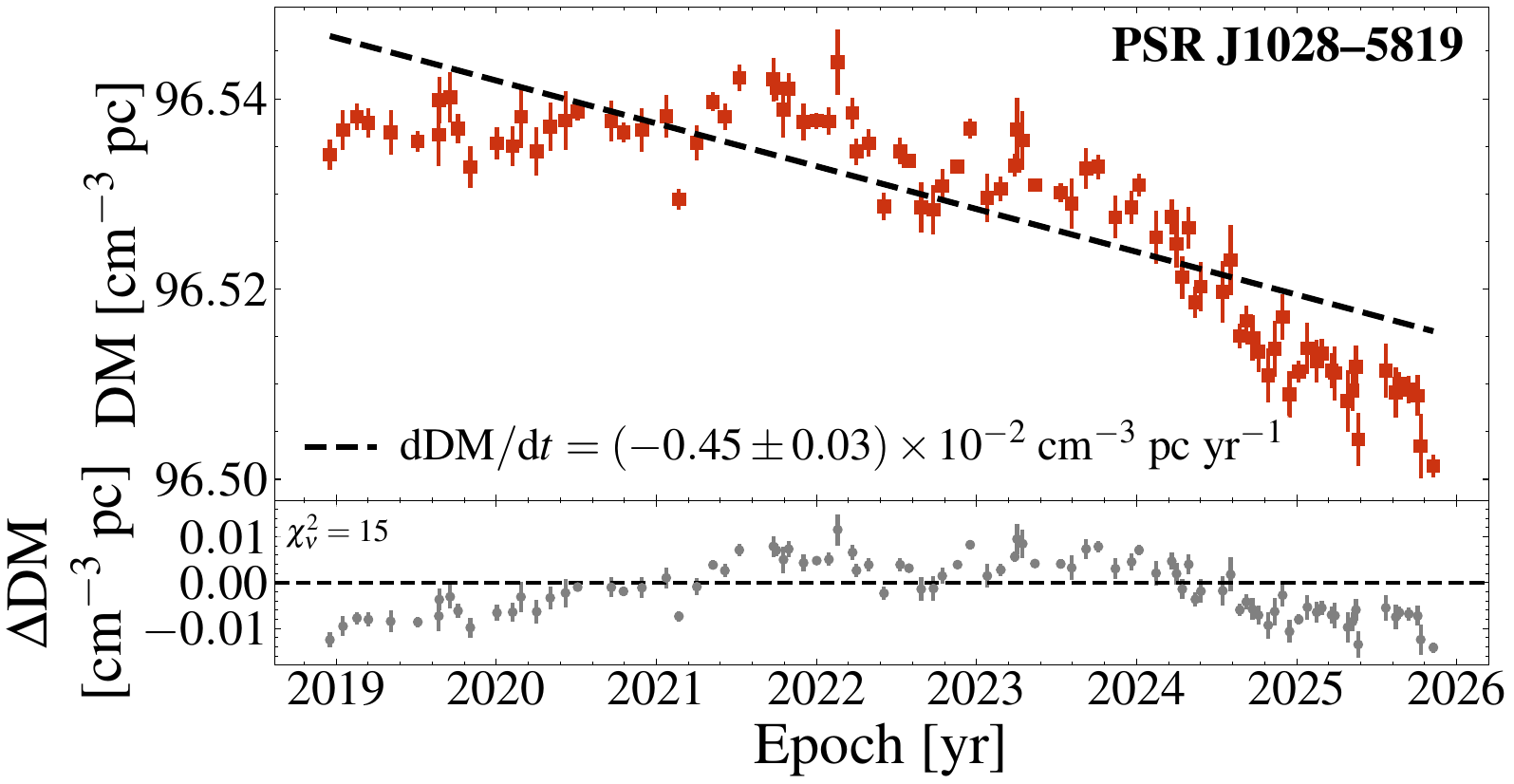}
    \caption{DM time series of PSR~J1028--5819. The red squares show the measured DM values with their corresponding $1\sigma$ uncertainties for the UWL dataset. The gray points show the DM values after subtracting the fitted linear trend. The reduced chi-squared ($\chi_\nu^2$) was computed after removing the linear trend.}
    \label{fig:J1028-5819}
\end{figure}

PSR~J1028--5819 \citep{keith2008} is a 91~ms pulsar with a spin-down luminosity of $8.3 \times 10^{35}$~ergs~s$^{-1}$. It is associated with a PWN \citep{ackermann2011,hess2018}. X-ray confirmation of the region is inconclusive because of a bright contaminating star in the field \citep{mignani2012}. No independent distance measurement is available; the DM implies a distance between 2.3~kpc using the NE2025 model and 1.4~kpc using the YMW16 model. This makes PSR~J1028--5819 a useful probe of DM variations arising from a PWN-enriched LoS, complementing the SNR-dominated cases of PSRs~J0835--4510 and J0908--4913.

Prior to our observations, a significant DM gradient in this pulsar could not be detected: \citet{petroff2013} placed only an upper limit of $|\text{dDM/d}t_{\rm lim}| \leq 0.006$~cm$^{-3}$~pc~yr$^{-1}$ from multibeam data. The multibeam data in our dataset similarly show large per-epoch uncertainties arising from the low S/N and are therefore excluded from the analysis. Using the UWL data alone, we measure a significant DM slope of $\text{dDM/d}t = (-4.50 \pm 0.35) \times 10^{-3}$~cm$^{-3}$~pc~yr$^{-1}$, consistent with the Petroff upper limit. Based on the DM SF, we find a spectral index of $\beta = 3.4 \pm 0.2$, which is a flatter slope than expected from a Kolmogorov-like turbulence but still consistent within $5 \sigma$. However, because of the short time span of our data, we cannot draw a definitive conclusion about the turbulence along the LoS with the PWN.

\subsubsection{PSR J1105--6107 and PSR J1829--1751}\label{sec:J1105-6107_J1829-1751}
\begin{figure*}
    \centering
    \begin{subfigure}[t]{0.49\textwidth}
        \includegraphics[width=\textwidth]{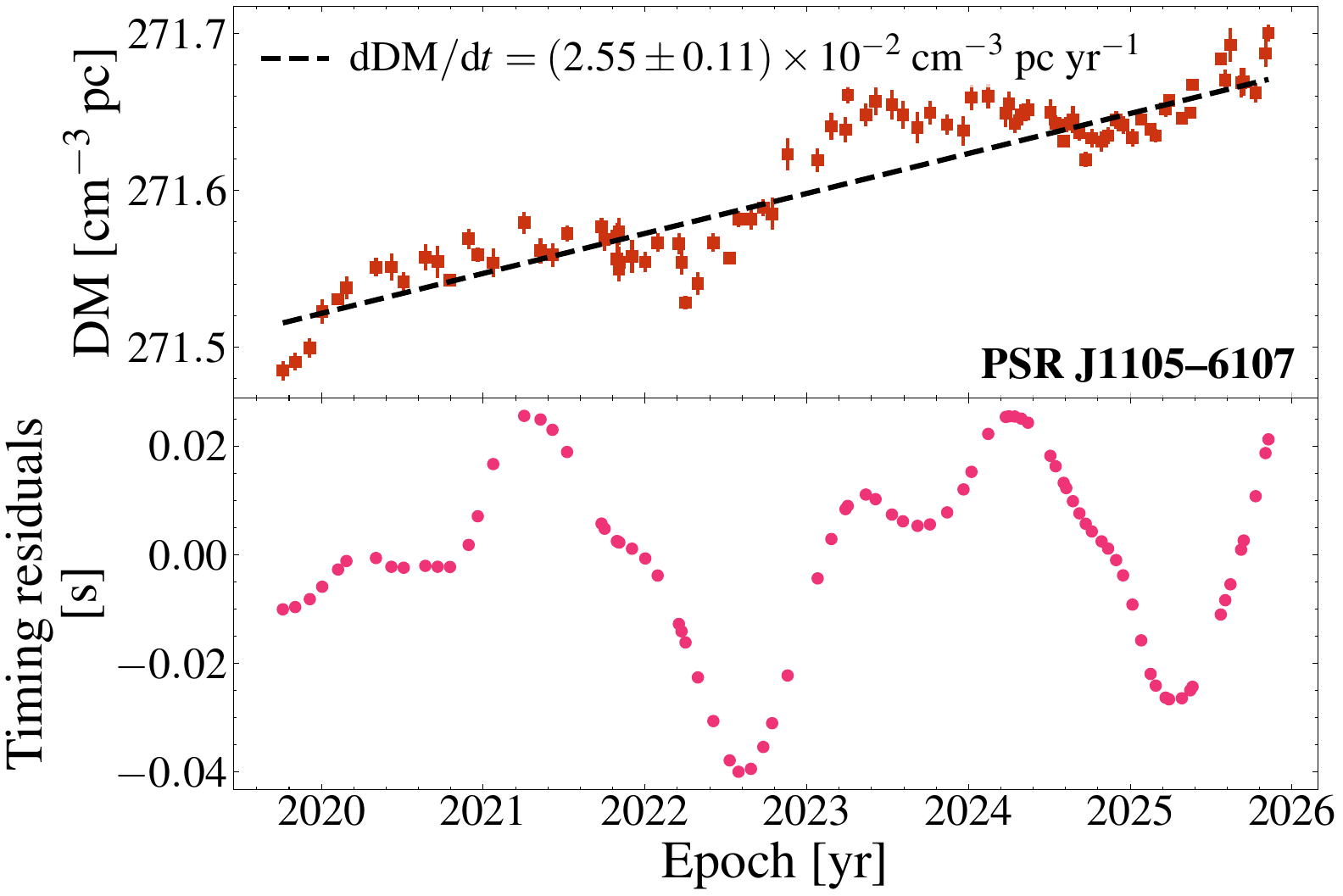}
    \end{subfigure}
    \hfill
    \begin{subfigure}[t]{0.49\textwidth}
        \includegraphics[width=\textwidth]{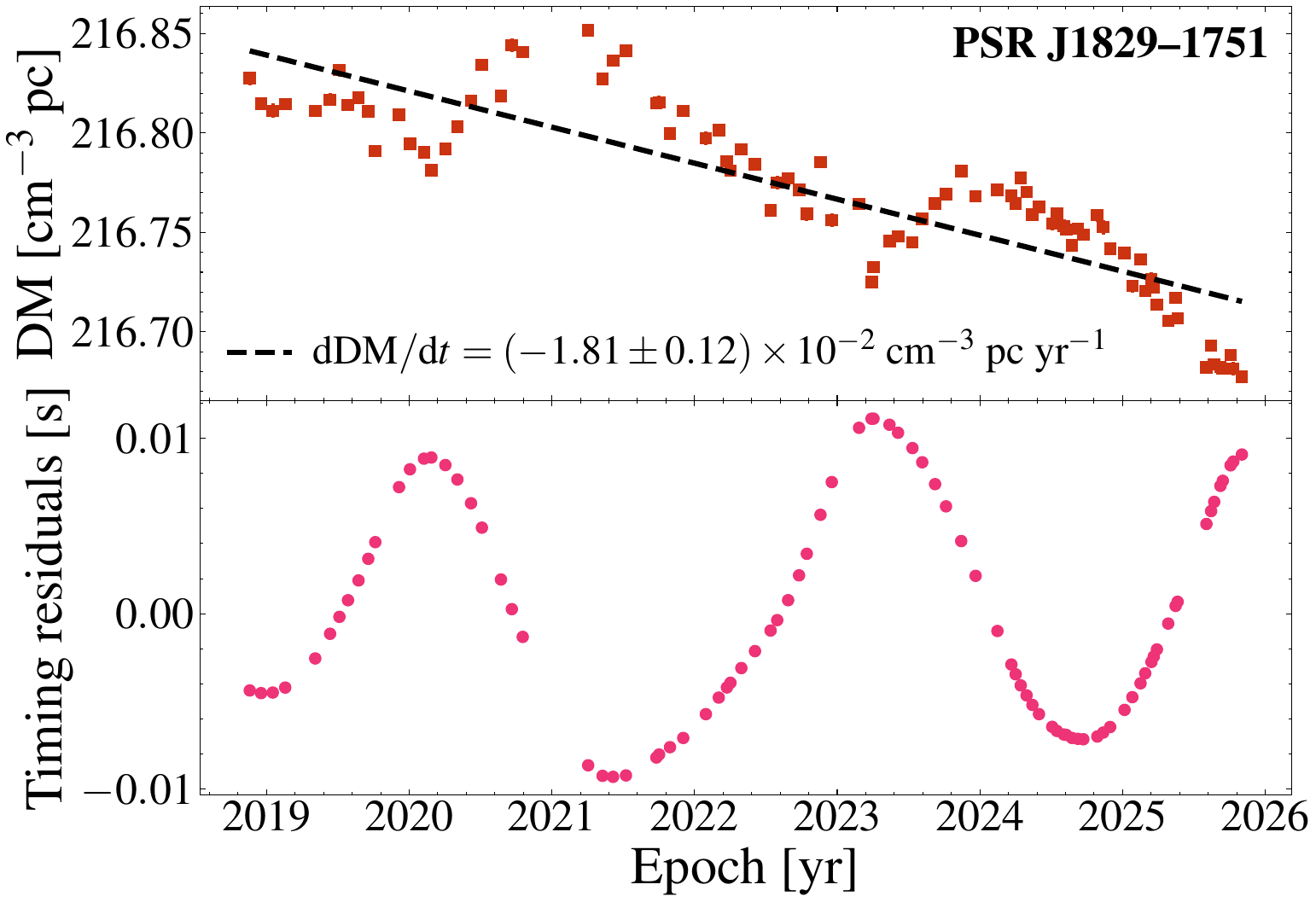}
    \end{subfigure}
    \caption{DM time series (\textit{top panels}) and post-fit timing residuals (\textit{bottom panels}) for PSRs~J1105--6107 (\textit{left}) and J1829--1751 (\textit{right}). The red squares show the measured DM values with their corresponding 1$\sigma$ uncertainties for the UWL dataset. The pink points show the post-fit timing residuals after subtracting the best-fitting timing model.}
\label{fig:J1105-6107_J1829-1751}
\end{figure*}

PSR~J1105--6107 \citep{kaspi1997} has a spin period of 63~ms and a spin-down energy loss rate of $2.5 \times 10^{36}$~ergs~s$^{-1}$. Upon its discovery, a possible association with the SNR G290.1--0.8 was proposed. Based on its DM value, the distance to this pulsar is estimated to be $\sim 2.4$ kpc using the YMW16 model. Compared to J1105--6107, PSR J1829--1751 \citep[B1826--17; ][]{davies1973} is a slower (300 ms) and less energetic ($\dot{E} = 7.6 \times 10^{33}$~ergs~s$^{-1}$) pulsar located at a DM-based distance between 4.4~kpc (NE2025) and 5.9~kpc (YMW16). The primary interest of these two pulsars is not in their local IISM environments, but in whether their measured DM variations genuinely trace IISM turbulence or are instead contaminated by pulse profile changes driven by their magnetospheres.

For both pulsars, we find a DM time series dominated by a linear trend with a superimposed quasi-periodic modulation. A Lomb-Scargle periodogram analysis shows a significant period of 0.347~yr$^{-1}$ ($\sim 1052$ days) for PSR~J1105--6107 and 0.329~yr$^{-1}$ ($\sim 1110$ days) for PSR~J1829--1751. The periodicity in PSR~J1829--1751 is not new: \citet{hobbs2010} identified a 2.4--3.2~yr periodicity in its timing residuals, and \citet{shaw2022}, \citet{nitu2022}, and \citet{lower2025} recovered the same period in its spin-down rate. PSR~J1105--6107 is also known to exhibit correlated profile and spin-down variations \citep{brook2016,lower2025}, although these are not strictly periodic. \citet{lyne2010} showed that such quasi-periodic modulations in young pulsars are driven by changes in the pulsar magnetosphere and are unrelated to their IISM environments. 

To test whether the same mechanism operates in PSR~J1105--6107, we compute the post-fit timing residuals of both pulsars using \textsc{tempo2}. The DM values (linear trend subtracted) and the timing residuals are shown in Fig.~\ref{fig:J1105-6107_J1829-1751}. For PSR~J1829--1751, we find a DM slope of $\text{dDM/d}t = (-1.81 \pm 0.12) \times 10^{-2}$~cm$^{-3}$~pc~yr$^{-1}$ over the UWL span and a Spearman correlation coefficient of $\rho = -0.636$ ($p$-value $3.18 \times 10^{-12}$) between the DM residuals and post-fit timing residuals. For PSR~J1105--6107, we find $\text{dDM/d}t = (2.55 \pm 0.11) \times 10^{-2}$~cm$^{-3}$~pc~yr$^{-1}$ and $\rho = 0.472$ ($p$-value $7.07 \times 10^{-6}$). Both correlations are significant. In contrast, PSR~J0908--4913, a pulsar known to show nonperiodic profile and spin-down variations \citep{brook2016,lower2025}, does not show significant correlation between the DM time series and the timing residuals ($\rho = 0.064$, $p$-value $0.73$) and does not exhibit any periodic structure in the DM variations. This strengthens the interpretation that the correlated behavior seen in PSR~J1829--1751 and PSR~J1105--6107 arises from periodic profile variability contaminating the DM measurements.

We emphasize that the apparent periodic DM variations observed in PSRs~J1105--6107 and J1829--1751 do not represent genuine changes in the electron density content along the LoS in the IISM. Instead, they arise because of the assumption of a static pulse template to model the frequency-dependent variations in the pulse profile that bias our sub-banded ToAs and, therefore, our DM measurements. The statistically significant correlations between the post-fit timing residuals and the DM variations support this hypothesis, which shows that the profile changes likely originating from the pulsar's magnetosphere introduce systematic biases. Disentangling these effects from other apparent periodic DM variations due to, for example, an intervening scattering screen requires a more detailed study into the frequency- and time-dependent evolution of the pulse profile using wide-band receivers such as the UWL receiver. 

\subsubsection{PSR J1359--6038}\label{sec:J1359-6038}
\begin{figure}
   \centering
    \begin{subfigure}[t]{\hsize}
        \includegraphics[width=\textwidth]{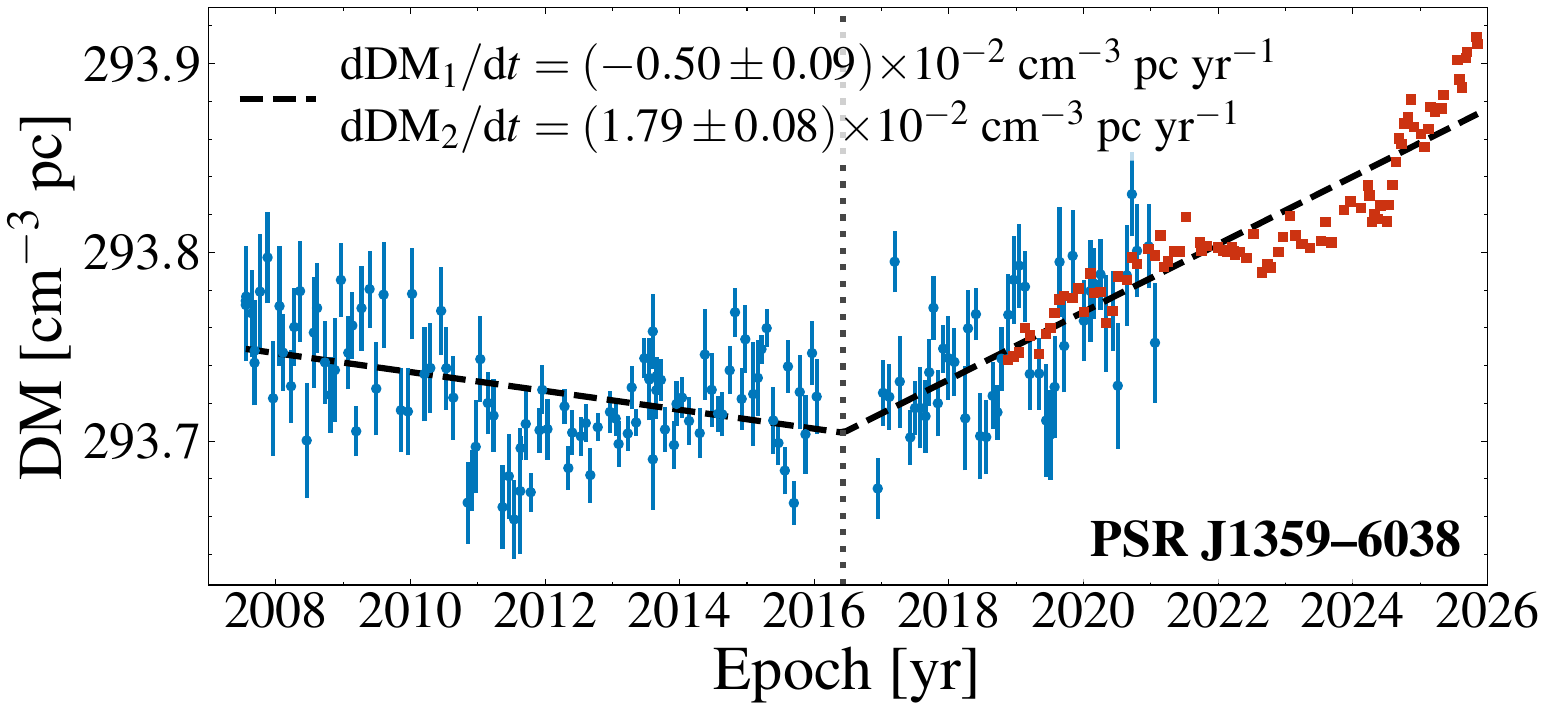}
    \end{subfigure}
    \vfill
    \vspace{0.2em}
    \begin{subfigure}[t]{\hsize}
        \includegraphics[width=\textwidth]{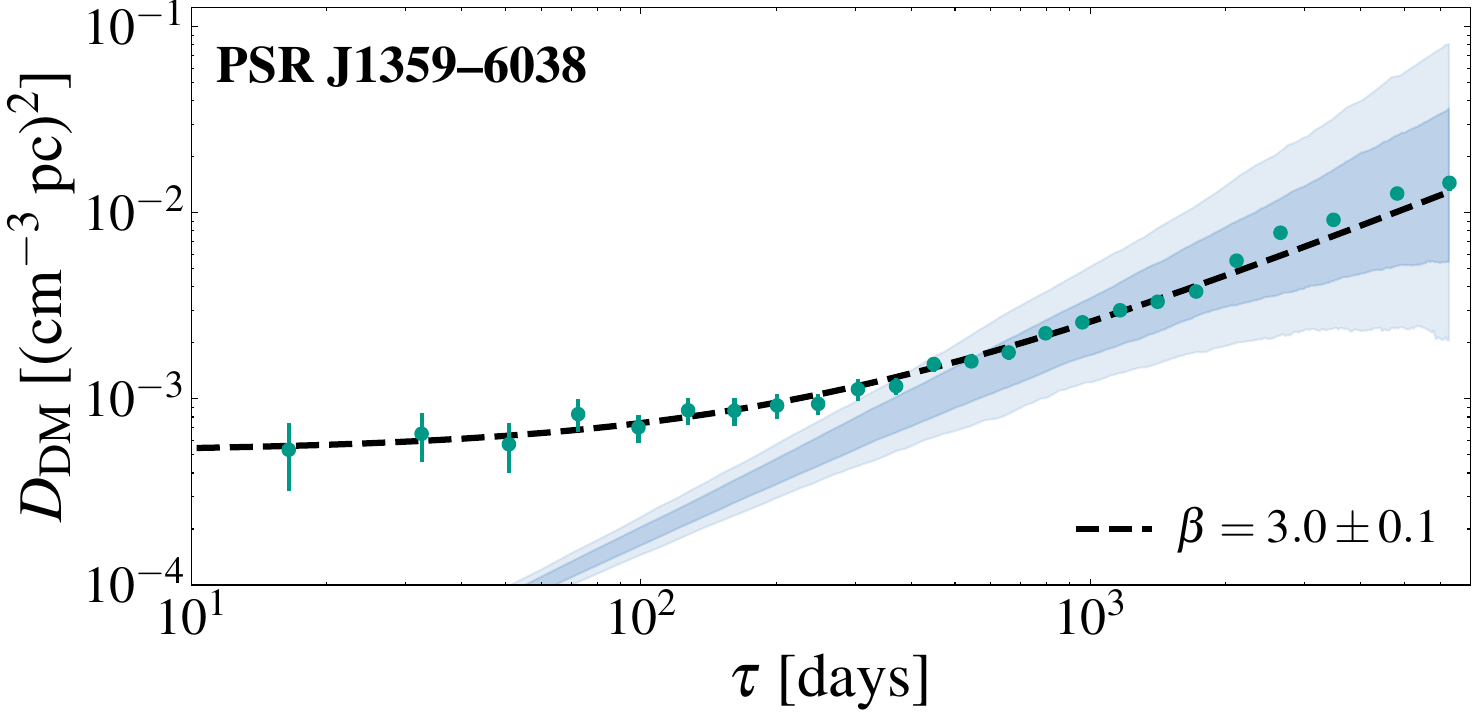}
    \end{subfigure}
    \caption{\textit{Top}: DM time series for PSR~J1359--6038. The points show the measured DM values with their corresponding $1\sigma$ uncertainties for the multibeam (blue) and UWL (red) datasets. The dashed black line shows the linear fit to multibeam data and the UWL data, fitted with a turnover indicated by the dotted vertical gray line. \textit{Bottom}: Same as the bottom panel of Fig.~\ref{fig:J0835-4510_J0908-4913} but for PSR~J1359--6038.}
    \label{fig:J1359-6038}
\end{figure}

PSR J1359--6038 \citep[B1356--60; ][]{manchester1978} is a pulsar with a period of 128 ms and a spin-down luminosity of $1.2 \times 10^{35}$~ergs~s$^{-1}$. Both DM-based distance measurements and work by \citet{verbiest2012} place the distance to the pulsar at $\sim 5$ kpc. Near this pulsar, \citet{gaensler2000} found a faint shell of radio emission G311.28+1.09 that could be an SNR. 

Figure~\ref{fig:J1359-6038} shows the DM time series for this pulsar for the multibeam and UWL datasets. We fit two linear trends around a change point to characterize the changing slope. Before June 2016, we find a negative DM gradient of $(-0.50\pm0.09) \times 10^{-2} \text{ cm}^{-3}\text{ pc yr}^{-1}$. Thereafter, we find a slope of $(1.79\pm0.08) \times 10^{-2} \text{ cm}^{-3}\text{ pc yr}^{-1}$, which is consistent with the findings of \citet{keith2024}. Based on the combined data, we find a wavenumber spectral slope of $\beta = 3.1 \pm 0.2$, in agreement with expectations for a Kolmogorov turbulence within $5\sigma$, although not very constraining. Although the SNR candidate and pulsar association remain uncertain, the observed DM variations could also arise from turbulence in the general IISM along the LoS.

\subsubsection{PSR J1413--6141}\label{sec:J1413-6141}
\begin{figure}
    \centering
    \begin{subfigure}[t]{\hsize}
        \includegraphics[width=\textwidth]{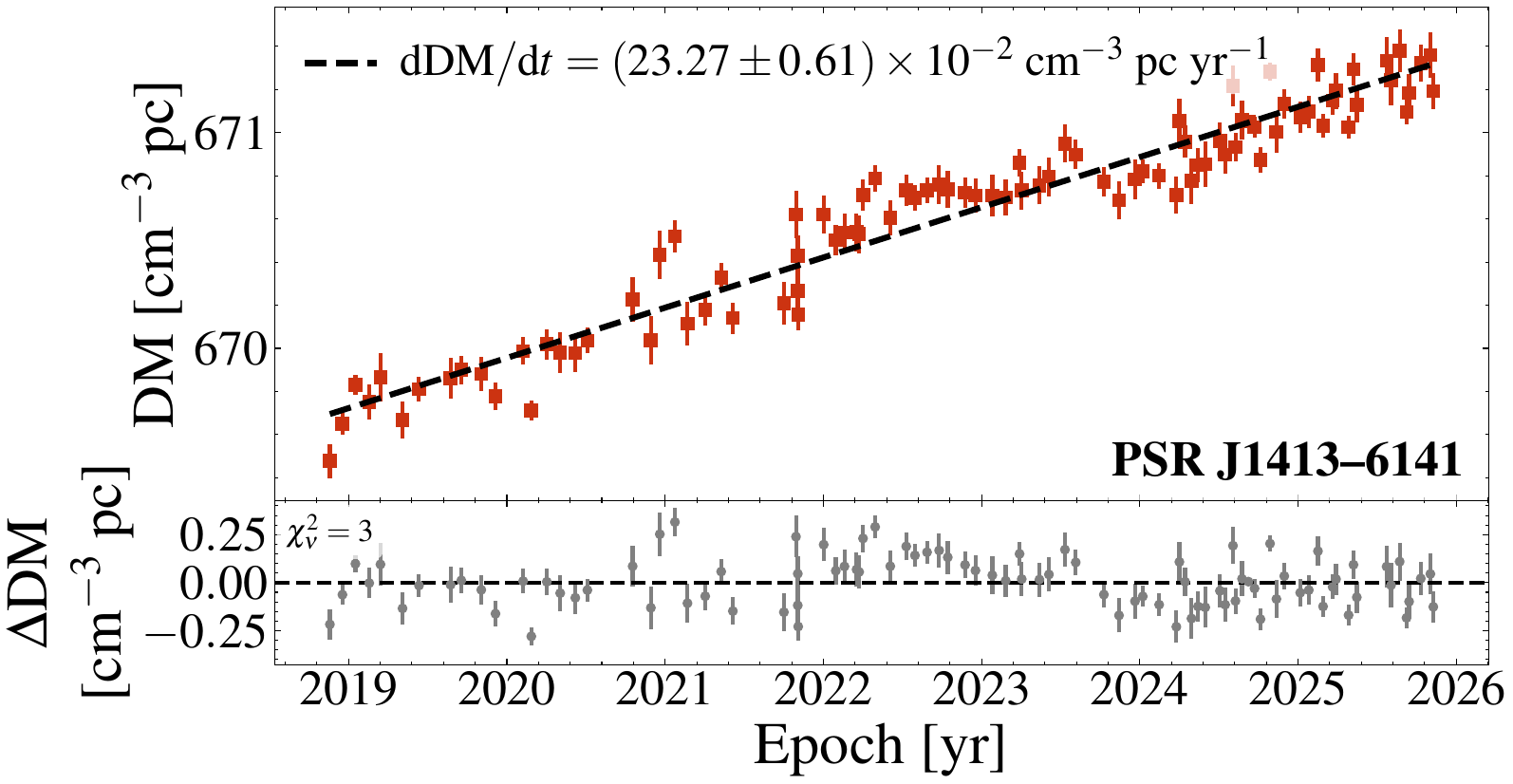}
    \end{subfigure}
    \caption{Same as Fig.~\ref{fig:J1028-5819} but for the DM time series of PSR~J1413--6141 from the UWL data.}
    \label{fig:J1413-6141}
\end{figure}

PSR~J1413--6141 \citep{kramer2003} has a spin period of 285~ms and a spin-down energy loss rate of $5.6 \times 10^{35}$~ergs~s$^{-1}$. It sits in one of the most complex regions of our sample: within $1^\circ$ on the sky, there are three additional pulsars of various ages \citep[e.g.,][]{chambery2023}, the SNR G312.4--0.4, and the gamma-ray source 3EG~J1410--6147. Early work by \citet{manchester2002} proposed an association between the SNR and PSR~J1413--6141, but \citet{doherty2003} subsequently found PSR~J1412--6145 to be a more likely link with the SNR. The gamma-ray emission is associated with a highly energetic pulsar J1410--6132 \citep{obrien2008}. PSR~J1413--6141 itself does not have confirmed gamma-ray emission \citep{smith2019}. Nevertheless, with a DM-based distance estimate between 8.5~kpc and 10~kpc, its LoS would still traverse material associated with both the SNR and the gamma-ray region, so these structures may still influence the observed DM variations even without a physical association.

PSR~J1413--6141 shows one of the steepest DM gradients in our sample. From the UWL data (Fig.~\ref{fig:J1413-6141}), we measure a slope of $(23.27 \pm 0.61) \times 10^{-2}$~cm$^{-3}$~pc~yr$^{-1}$. The DM time series is strongly dominated by this linear trend, which is reflected in the spectral slope value of $\beta = 4.0 \pm 0.5$ in the SF. 

The physical origin of this gradient is difficult to isolate. If the dominant contribution is the pulsar's transverse motion sweeping through an electron density gradient, the implied mean electron density along the LoS would be $n_e \gtrsim1.1 \times 10^4$~cm$^{-3}$, assuming the transverse velocity upper limit of 200~km~s$^{-1}$ from \citet{doherty2003}. This density is unphysically high for a typical IISM environment, exceeding even dense H\,{\sc ii} regions by an order of magnitude, which suggests that transverse motion alone cannot account for the observed slope. An additional contribution from the pulsar's radial motion is likely: as the pulsar moves along the LoS, its distance to Earth changes gradually, altering the total integrated DM even in a uniform medium \citep[e.g.,][]{lam2016}. The steep gradient therefore most probably reflects a combination of radial motion and transverse sampling of dense structures along the LoS (such as the SNR G312.4--0.4 material). Precise measurements of the proper motion of this pulsar, as well as distance estimates, could help disentangle the relative contributions to the observed DM variations.

\subsubsection{PSR J1512--5759}\label{sec:J1512-5759}

\begin{figure}
    \centering
    \begin{subfigure}[t]{\hsize}
        \includegraphics[width=\textwidth]{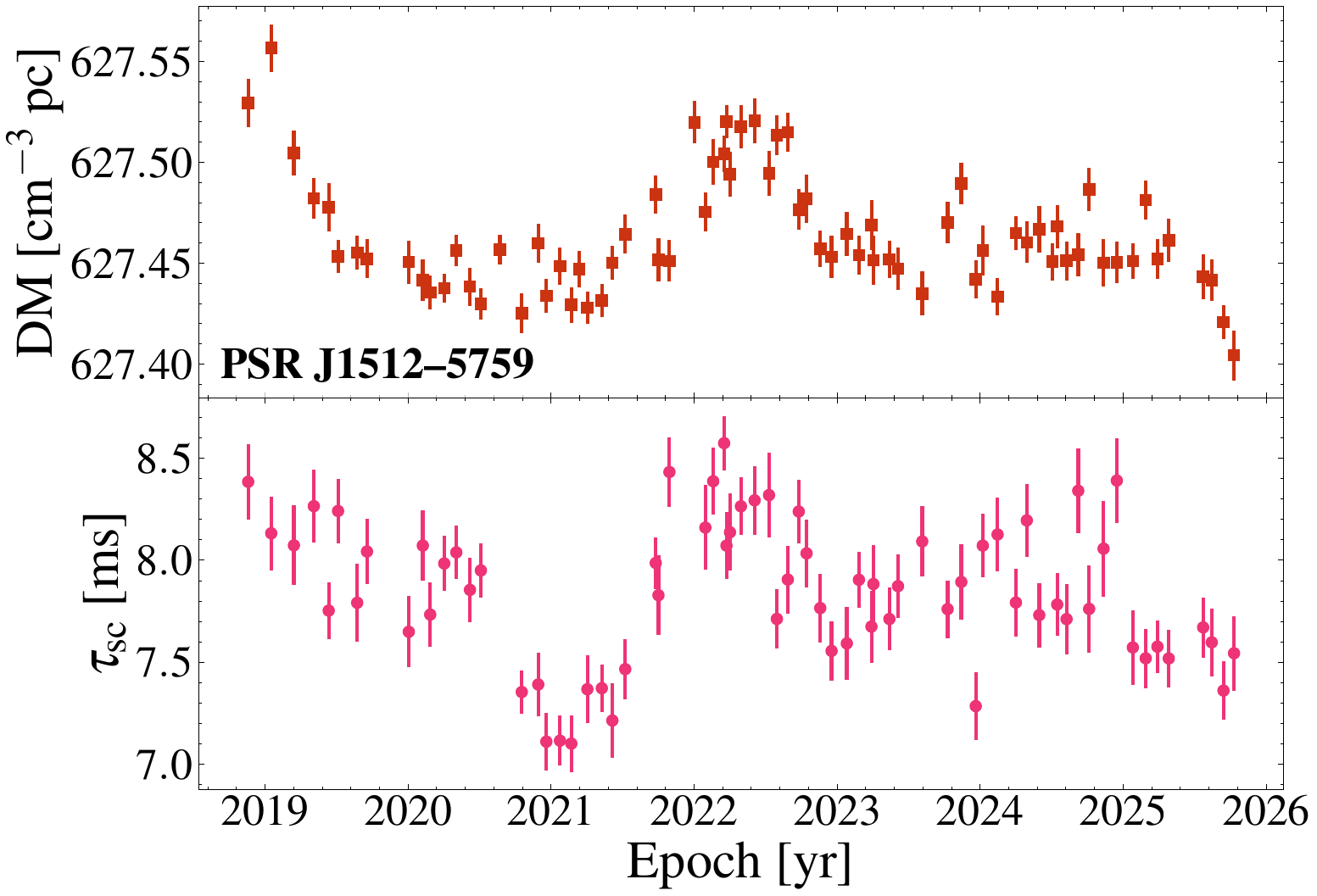}
    \end{subfigure}
    \caption{Time series of chromatic timing delays for PSR~J1512--5759. \textit{Top}: Dispersive delay time series. The red squares show the measured DM values with their corresponding 1$\sigma$ uncertainties for the UWL dataset. \textit{Bottom}: Scattering delay time series. The pink points show the scattering timescale at 1 GHz inferred from a direct fit to the frequency-dependent profiles.}
    \label{fig:J1512-4749}
\end{figure}

PSR~J1512--5759 \citep[B1508--57; ][]{johnston1992} is a 128 ms period pulsar with a spin-down luminosity of $1.3 \times 10^{35}$~ergs~s$^{-1}$. The dispersive delay variations of PSR~J1512--5759 are not captured by the parameterization of a linear gradient, as can be seen in Fig.~\ref{fig:J1512-4749}, which highlights the limitation of using a DM gradient to parameterize the DM variations. Moreover, the pulse profile is also visibly scattered. This makes this pulsar an ideal test case to study the correlations between the DM and scattering.

Figure~\ref{fig:J1512-4749} shows the DM and scattering timescale at 1 GHz as a function of time for this pulsar. There is a clear correlation between the dispersive and scattering delays with a Spearman coefficient of $\rho = 0.493$ ($p$-value $1.66 \times 10^{-5}$). Additionally, we find an average scattering index of $\alpha = 3.21 \pm 0.16$. Although this value is inconsistent with a simple scattering screen or a Kolmogorov turbulence, both MSPs and canonical pulsars are known to deviate from these values \citep[e.g.,][]{geyer2017,krishnakumar2019,oswald2021}. 

We cannot exclude leakage from the scattering delay in our DM measurements. Moreover, both effects could arise from the same ionized structures in the IISM. Density variations in our LoS simultaneously contributed to the integrated electron column density that is responsible for the dispersive delay, as well as the small-scale fluctuations that are responsible for the multipath propagation of the radio waves. This makes it difficult to disentangle whether the observed epoch-to-epoch variations are primarily driven by the variations in the DM, scattering, or a combination of the two. Advanced profile modeling techniques, such as \textsc{PulsePortraiture} \citep{pennucci2016}, could help differentiate between these effects.

\subsubsection{PSR J1731--4744}
PSR~J1731--4744 \citep[B1727--47;][]{large1968} is a pulsar with a spin period of 830 ms and a spin-down energy loss rate of $1.1 \times 10^{34}$~ergs~s$^{-1}$. The DM-based distance estimate places the pulsar at approximately 2.2 kpc (NE2025) and 5.5 kpc (YMW16) from Earth. \citet{shternin2019} show that the pulsar's proper motion is consistent with originating from the SNR G343.0--6.0 (RCW 114). If this association is correct, the pulsar would instead lie at a much smaller distance of $\sim 0.6$ kpc. This possible association is further supported by the results of \citet{main2023} and \citet{lower2021}.

\begin{figure}
    \centering
    \begin{subfigure}[t]{\hsize}
        \includegraphics[width=\textwidth]{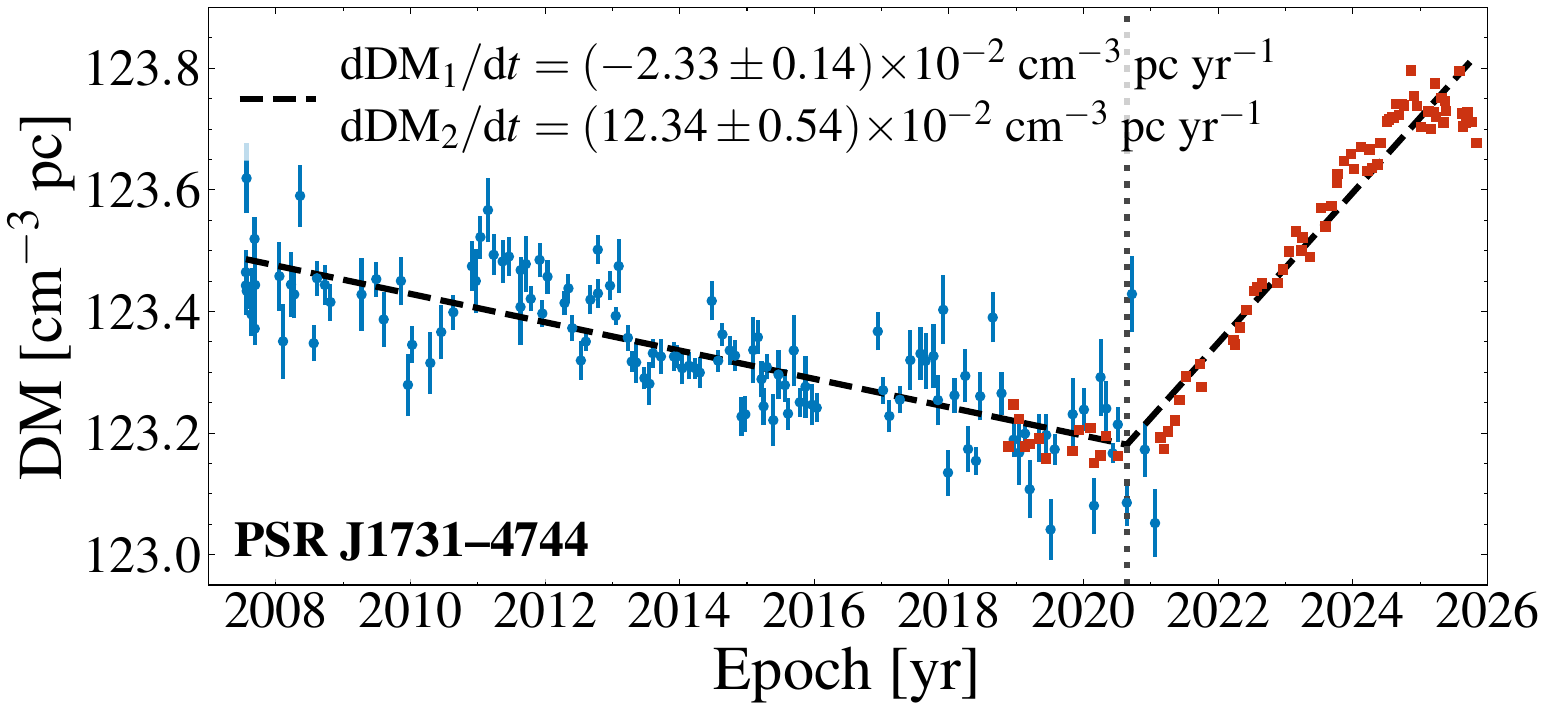}
    \end{subfigure}
    \vfill
    \vspace{0.2em}
    \begin{subfigure}[t]{\hsize}
        \includegraphics[width=\textwidth]{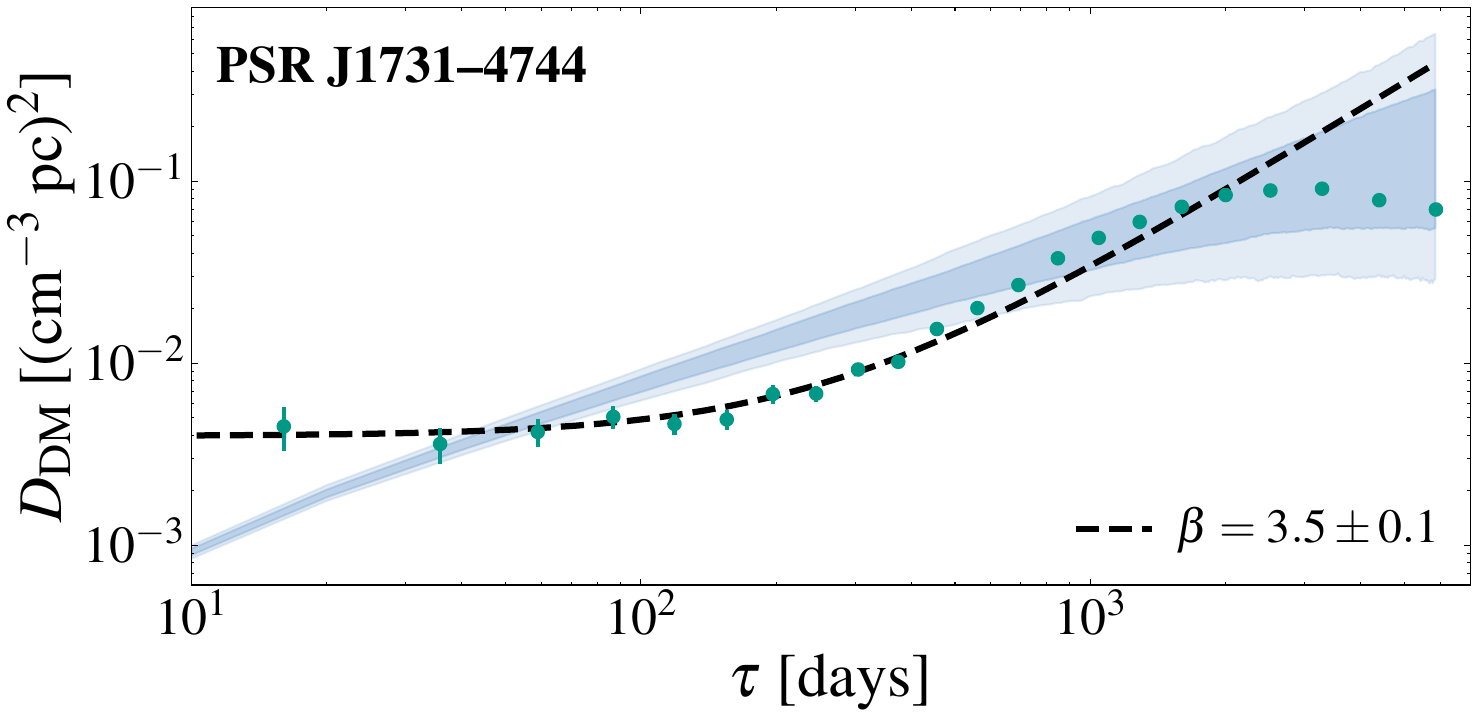}
    \end{subfigure}
    \caption{Same as Fig.~\ref{fig:J1359-6038} but for PSR~J1731--4744.}
    \label{fig:J1731-4744}
\end{figure}

The DM variations observed in PSR J1731--4744 are among the most extreme in our sample (see Fig.~\ref{fig:J1731-4744}). Before August 2020, the data show a negative slope of $(-2.33\pm 0.14) \times 10^{-2}$~cm$^{-3}$~pc~yr$^{-1}$. After this, the trend increases sharply with a positive slope of $(12.34\pm 0.54) \times 10^{-2}$~cm$^{-3}$~pc~yr$^{-1}$. Moreover, the rotation measure time series exhibits an identical correlation compared to the DM time series \citep[][Melfor et al. in prep.]{johnston2021}, supporting the fact that the electron column density is changing in the LoS.

The DM SF is not well described by a simple power law with a constant noise floor and therefore appears inconsistent with Kolmogorov turbulence over the sampled length scales. However, this does not necessarily imply that the underlying wavenumber spectrum deviates from a Kolmogorov form. Because the observations sample only a finite range of spatial scales from a single realization of the turbulent medium, the measured SF could reflect a portion of the spectrum near a turnover or break. In this case, the apparent deviation from a power law would arise from the limited data span rather than from a fundamentally different turbulence spectrum in the IISM.

Alternatively, the observed DM variations could be influenced by the local environment of the associated SNR. This interpretation is supported by the results of \citet{main2023}, who measured a short scintillation timescale and clearly resolved scintles in frequency, indicating a scattering screen located close to the pulsar. While some pulsars embedded in or close to SNRs, such as PSR~J0835--4510 and PSR~J0908--4913, exhibit DM SFs consistent with Kolmogorov turbulence, others show more complex behavior.

Distinguishing between these possibilities requires continued long-term monitoring of the DM variations of PSR~J1731--4744. In addition, measurements of the DISS timescale and bandwidth over a longer baseline would provide complementary constraints on the scattering environment and help determine whether the observed variations are consistent with Kolmogorov turbulence or instead arise from localized plasma structures.

\subsubsection{PSR J1833--0827}\label{sec:J1833-0827}

\begin{figure}
   \centering
   \begin{subfigure}[t]{\hsize}
        \includegraphics[width=\textwidth]{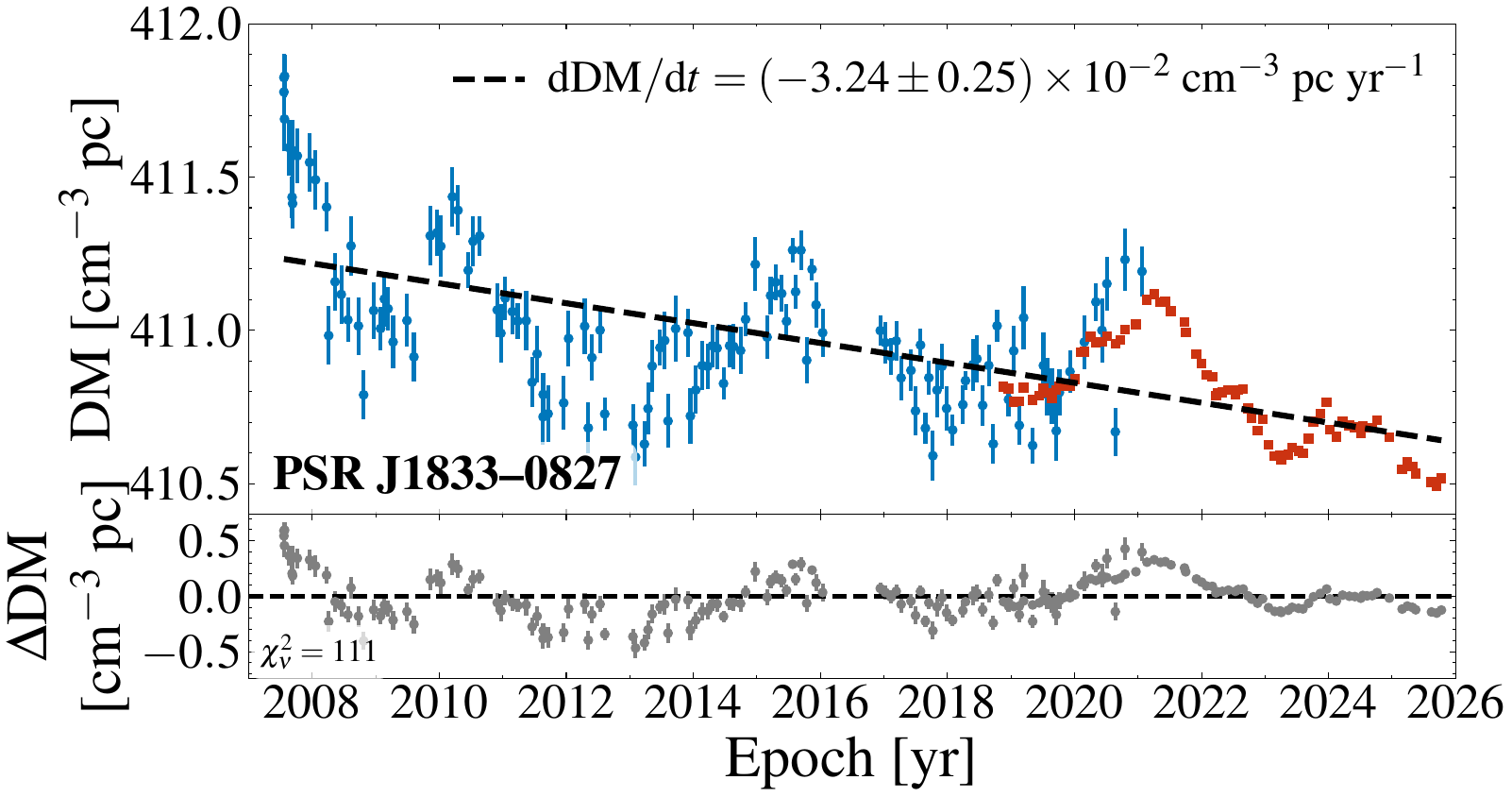}
    \end{subfigure}
    \vfill
    \vspace{0.2em}
    \begin{subfigure}[t]{\hsize}
        \includegraphics[width=\textwidth]{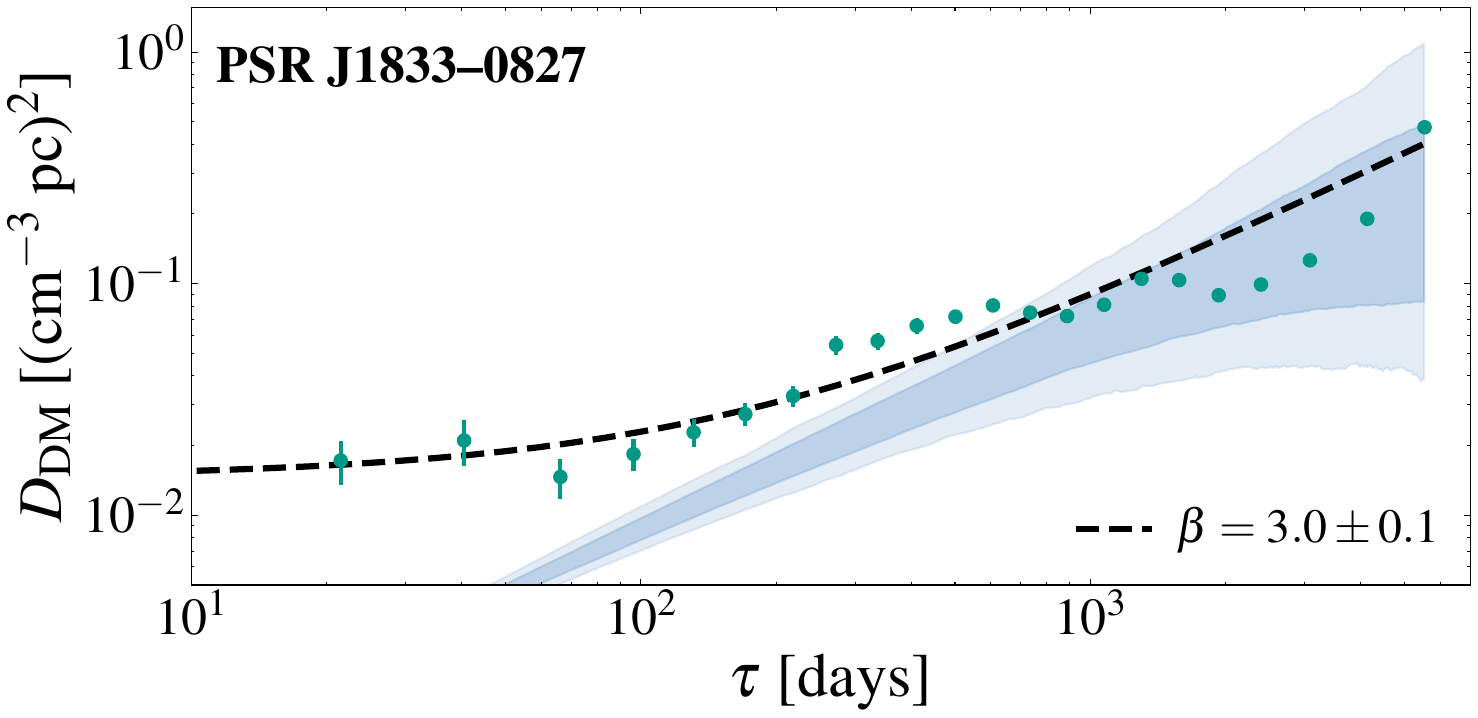}
    \end{subfigure}
    \caption{Same as Fig.~\ref{fig:J0835-4510_J0908-4913} but for PSR~J1833--0827.}
    \label{fig:J1833-0827}
\end{figure}

PSR~J1833--0827 \citep[B1830--08;][]{clifton1986} is a pulsar with a spin period of 85 ms and a spin-down energy loss rate of $5.8 \times 10^{35}$~ergs~s$^{-1}$. The pulsar lies close to the radio-bright shell SNR G23.3--03. A proper motion measurement by \citet{hobbs2005} is consistent with the pulsar originating within this remnant. In addition, a relic PWN associated with the pulsar was detected by \citet{esposito2011} and later confirmed by \citet{kargaltsev2012} using XMM-Newton observations.

Figure~\ref{fig:J1833-0827} shows the DM time series for PSR~J1833--0827 derived from the multibeam and UWL datasets. We measure an overall decreasing slope of $(-3.24\pm 0.25) \times 10^{-2}$~cm$^{-3}$~pc~yr$^{-1}$. Although \citet{keith2024} reported a positive slope, their result reflects the increase in DM observed between approximately 2020 and 2022. Earlier work by \citet{petroff2013} reported a strong decreasing slope of $-0.13 \pm 0.02$~cm$^{-3}$~pc~yr$^{-1}$, the largest DM gradient detected in their sample.

\citet{petroff2013} suggested that these DM variations could originate from the X-ray PWN, with a dense filament crossing the LoS when the DM increases. However, in our data, the DM variations appear to show a quasi-periodic structure. Such behavior is difficult to reconcile with a single transient filament crossing the LoS and instead suggests a more persistent or structured origin for the observed variations in the time series.

Fitting a power-law model with a constant noise floor to the DM SF yields a wavenumber spectral index of $\beta = 3.0 \pm 0.1$, significantly shallower than expected for a Kolmogorov turbulence. Moreover, the measured SF shows larger deviations from a simple power-law behavior. 

Assuming a DM-based distance of 4.38 kpc, \citet{parthasarathy2020} derived a transverse velocity of $800 \pm 200$~km~s$^{-1}$, consistent with the proper motion measurements reported by \citet{hobbs2004}. Using Eq.~\ref{eq:cn2_approx}, we estimate a spectral coefficient of $C_n^2 \simeq 10^{-1.3} \text{ m}^{-20/3}$. This value is comparable to that obtained for PSR~J0908--4913, although we emphasize that this estimate assumes that the observed turbulence follows a Kolmogorov spectrum. The enhanced $C_n^2$ in combination with the shallower-than-expected value for the slope suggests that the DM variations along the LoS are unlikely to arise from a uniform Kolmogorov medium. A possible explanation for the observed quasi-periodic DM variations is that the LoS crosses one or more discrete, slab-like electron-density structures \citep{lam2016}. Depending on their orientation relative to the LoS and the pulsar velocity, the LoS intersection samples different slab-like structures over time (see, e.g., Fig.~9 of \citealt{lam2016}), which can cause the semi-periodic trend observed in the DM time series. Moreover, multiple filaments associated with the surrounding PWN could plausibly give rise to the repeated, quasi-periodic DM variations we observe. 

Nevertheless, the inferred value of $C_n^2$ suggests enhanced density fluctuations along the LoS. A more detailed investigation of the local IISM toward PSR~J1833--0827 will be required to determine the origin of the observed periodicity in the DM variations.

\section{Conclusions}\label{sec:conclusions}
We have analyzed DM variations in a sample of 364 young, energetic pulsars observed with the Murriyang/Parkes radio telescope over 18 years, combining legacy multibeam data with UWL observations. The substantially wider bandwidth and improved per-epoch DM precision of the UWL receiver, combined with the extended time baseline, have enabled the most sensitive characterization of DM variability in young pulsars to date.

We detect significant ($\geq 5\sigma$) DM gradients in 23 and 89 pulsars in the multibeam and UWL datasets, respectively, with 16 pulsars in the combined dataset showing significant variations. This represents a substantial improvement over the four detections reported by \citet{petroff2013} from an overlapping sample, which we attribute to our longer time baseline and the order-of-magnitude improvement in DM precision provided by the UWL receiver. The measured DM gradients are consistent with the power-law $|{\rm dDM/d}t| \propto \mathrm{DM}^{0.81}$ relation, which is steeper than the relation of \citet{hobbs2004}, albeit within the error bars of their inferred slope. We attribute this to the fact that only 21 out of their 374 pulsars show a significant DM slope above $3\sigma$ \citep{donner2020}. Moreover, the DM fluctuations have a dependence on the pulsar's velocity, and the resulting spread in velocities contributes to a corresponding spread in the measured DM slopes. This makes it difficult to clearly distinguish between the different scaling relations.

The majority of our pulsars with detected DM variations are consistent with a Kolmogorov wavenumber spectrum ($\beta = 11/3$) within $5\sigma$: out of 38 pulsars in the UWL dataset with a constrained $\beta$ in the Bayesian fit, 31 are consistent with a Kolmogorov spectral index. For the combined dataset, this is 5 out of 6 pulsars. However, due to the uncertainties on $\beta$, a substantial fraction of these are also consistent with $\beta = 4$, the value expected when the DM SF is dominated by a linear DM gradient: this is the case for 28 of the 38 UWL pulsars and 5 of the 6 combined dataset pulsars. We note that we did not model the contribution to the SF that arises from the linear gradient itself, which introduces a bias in $\beta$ inferred from the SF slope. Nevertheless, the larger fraction of sources consistent with $\beta = 11/3$ than with $\beta = 4$ provides tentative evidence that the observed DM variations are predominantly consistent with Kolmogorov turbulence. We highlight several pulsars with notable DM variations that show the influence of local environments on the observed turbulence:

PSR~J0835--4510 and PSR~J0908--4913, both associated with SNRs and PWNe, show DM SFs that are consistent with Kolmogorov turbulence within $3 \sigma$. However, they have spectral coefficients of $C_n^2 \simeq 10^{0.4}$~m$^{-20/3}$ and $C_n^2 \simeq 10^{-2.4}$~m$^{-20/3}$, respectively, which are significantly higher than expected from the diffuse IISM. This demonstrates that turbulent energy injected by the SNR cascades to large scales in a manner that preserves the Kolmogorov spectral shape, with the SNR environment increasing only the amplitude of the turbulent fluctuations.

Several other LoSs exhibit DM variations that cannot be reconciled with a broad picture of Kolmogorov turbulence in the IISM. PSR~J1413--6141 shows an exceptionally high DM gradient, which implies an unphysically high electron density if attributed to the transverse motion of the pulsar alone. PSR~J1731--4744 and PSR~J1833--0827 display a sign reversal in the slope, non-Kolmogorov SFs, and quasi-periodic behavior in the DM variations.

PSR~J1512--5759 shows that DM and scattering delays can be strongly correlated, which either indicates that the variations originate from the same turbulent structures in the LoS or that our DM measurements are biased because of the leakage of the scattering delays into our estimates. On the other hand, PSR~J1105--6107 and PSR~J1829--1751 demonstrate that not all apparent DM variations are due to an electron-density gradient in the LoS and that frequency-dependent profile evolution can mimic IISM propagation effects.

A concerning limitation encountered in this analysis is the contamination of wideband DM measurements by pulse jitter noise. We find this for 21 bright, long-period pulsars. The wide fractional bandwidth of the UWL, which reduces the formal DM uncertainty by an order of magnitude relative to the multibeam receiver, simultaneously increases sensitivity to intrinsic pulse-to-pulse profile variability. The jitter-induced epoch-to-epoch DM scatter exceeds the contribution expected from IISM turbulence, rendering the DM time series for these pulsars unreliable as a probe of the IISM. This issue will likely become even more pronounced with next-generation facilities such as the SKA. The increased sensitivity is expected to enable formal DM precisions as low as $\sim 10^{-8}$~cm$^{-3}$~pc \citep{tiburzi2025} and yield extremely high-S/N measurements of many pulsars. In practice, this is impossible to achieve, as jitter noise, DISS, and profile evolution will remain a fundamental floor for these sources. The SKA will be a heavily oversubscribed facility, so it will be difficult to obtain the required longer per-pulsar integration times for a large sample of pulsars such as ours. This will be especially problematic for IISM studies because the most informative targets are young pulsars, which are located in highly turbulent IISM environments. At the same time, their periods are often longer than those of MSPs and, therefore, are more susceptible to jitter on short integration times. Without a robust strategy to mitigate jitter, the next generation of radio telescopes may offer limited advantages over current instruments such as the Murriyang/Parkes radio telescope for IISM studies of these sources. A dedicated strategy to separate jitter-dominated DM measurements from genuine IISM variability will be essential to fully exploit the sensitivity of next-generation instruments.

Taken together, our results provide tentative evidence for the long-standing picture of a universal Kolmogorov power-law wavenumber spectrum for electron density fluctuations in the IISM \citep[cf.][]{armstrong1995}, although the uncertainties on $\beta$ and our neglect of the linear-gradient contribution to the SF prevent us from coming to a definitive conclusion. Young pulsars offer a uniquely sensitive probe of this spectrum through their proximity to active SNRs and star-forming regions, which increase the turbulence amplitude while preserving the Kolmogorov spectral shape. The systematic deviations seen in a minority of LoSs suggest structured local environments that require multiwavelength follow-up to disentangle.

\begin{acknowledgements}
      Murriyang, CSIRO’s Parkes radio telescope, is part of the Australia Telescope National Facility (\href{https://ror.org/05qajvd42}{https://ror.org/05qajvd42}) which is funded by the Australian Government for operation as a National Facility managed by CSIRO. We acknowledge the Wiradjuri people as the Traditional Owners of the Observatory site. S.J.I.M., A.P., and A.C. acknowledge financial support from the European Research Council (ERC) starting grant 'GIGA' (grant agreement number: 101116134). A.P. also acknowledges support through the NWO-I Veni fellowship. A.C. acknowledges financial support from the "Programme National de Cosmologie et Galaxies" (PNCG), "Programme National Hautes Energies" (PNHE) and "Programme National Gravitation, Références, Astronomie, Métrologie" (PNGRAM) of CNRS/INSU, France. A.C. acknowledges financial support from Agence Nationale de la Recherche (ANR-18-CE31-0015), France. Work at NRL is supported by NASA. M.E.L. is supported by the Australian Research Council (ARC) Discovery Early Career Research Award DE250100508. L.S.O. acknowledges support from the EPSRC Stephen Hawking Fellowship grant EP/Z534730/1. Part of this work was undertaken with support from the ARC Centre of Excellence for Gravitational Wave Discovery (OzGrav; CE230100016). S.J.I.M. thanks Joris Verbiest for insightful discussions on the IISM. We thank the referee, Dr. Michael Lam, for his insightful comments that improved the quality of the manuscript.
\end{acknowledgements}

\bibliographystyle{aa}
\bibliography{bib}

\begin{appendix}
\onecolumn
\section{Pulsars with significant DM variations}\label{app:significant_pulsars}

\begin{longtable}{lcccccc}
\caption{Pulsars with significant multibeam DM slopes.}\\
\hline
\hline
PSR & DM & (dDM/d$t$)$_{\rm multibeam}$ & $\beta_{\textsc{bilby}}$ & $D_{\rm DM,\,\textsc{bilby}}(1000\,\mathrm{d})$ & $\beta_{\rm direct}$ & $D_{\rm DM,\,\rm direct}(1000\,\mathrm{d})$ \\
& [cm$^{-3}$ pc] & [$\times\, 10^{-3}$ cm$^{-3}$ pc yr$^{-1}$] & & [$\times\, 10^{-3}$ (cm$^{-3}$ pc)$^2$] & & [$\times\, 10^{-3}$ (cm$^{-3}$ pc)$^2$] \\
\hline
\endfirsthead
\caption{continued}\\
\hline
\hline
PSR & DM & (dDM/d$t$)$_{\rm multibeam}$ & $\beta_{\textsc{bilby}}$ & $D_{\rm DM,\,\textsc{bilby}}(1000\,\mathrm{d})$ & $\beta_{\rm direct}$ & $D_{\rm DM,\,\rm direct}(1000\,\mathrm{d})$ \\
& [cm$^{-3}$ pc] & [$\times\, 10^{-3}$ cm$^{-3}$ pc yr$^{-1}$] & & [$\times\, 10^{-3}$ (cm$^{-3}$ pc)$^2$] & & [$\times\, 10^{-3}$ (cm$^{-3}$ pc)$^2$] \\
\hline
\endhead
\hline
\endfoot
\hline
\noalign{\smallskip}
\multicolumn{7}{@{}p{\textwidth}@{}}{\footnotesize\textbf{Notes.} Pulsars whose measured DM slopes are inconsistent with zero at the $\geq 5\sigma$ level in the multibeam data. Each pulsar in the table lists the DM, the linear fit to the DM time series (dDM/d$t$)$_{\rm multibeam}$, and the DM SF characterization from two independent methods: the Bayesian (\textsc{bilby}) fit and the least-squares fit. For each method, we report the power-law wavenumber spectrum index $\beta$ inferred from the DM SF and the value of the DM SF at a time lag of 1000 days, $D_{\rm DM}(1000\,\mathrm{d})$. Pulsars whose $\beta$ is constrained using the Bayesian fit are shown in bold.} \\
\endlastfoot
J0536--7543 & $18.6$ & $38 \pm 4$ & $2^{+2}_{-2}$ & $580^{+450}_{-244}$ & $6 \pm 3$ & $592$ \\
J0630--2834 & $34.4$ & $47 \pm 8$ & $2^{+2}_{-2}$ & $875^{+659}_{-375}$ & $6 \pm 4$ & $878$ \\
\textbf{J0738--4042} & $160.9$ & $-23.3 \pm 0.3$ & $3.3^{+0.3}_{-0.3}$ & $10^{+3}_{-2}$ & $3.37 \pm 0.03$ & $11$ \\
J0742--2822 & $73.8$ & $1.4 \pm 0.2$ & $2^{+2}_{-2}$ & $0.56^{+0.05}_{-0.04}$ & $2.1 \pm 0.3$ & $0.64$ \\
\textbf{J0835--4510} & $67.8$ & $-13.9 \pm 0.2$ & $3.4^{+0.3}_{-0.2}$ & $3.5^{+0.8}_{-0.7}$ & $3.64 \pm 0.03$ & $3.4$ \\
\textbf{J0908--4913} & $180.2$ & $-21.0 \pm 0.1$ & $3.8^{+0.3}_{-0.3}$ & $4.2^{+1.0}_{-0.8}$ & $3.83 \pm 0.09$ & $4.3$ \\
J1016--5857 & $393.6$ & $-63 \pm 11$ & $2^{+2}_{-2}$ & $982^{+95}_{-84}$ & $2.8 \pm 0.7$ & $1057$ \\
J1048--5832 & $128.7$ & $-8.1 \pm 0.7$ & $2^{+2}_{-2}$ & $7.8^{+0.7}_{-0.6}$ & $2.9 \pm 0.3$ & $9.5$ \\
J1105--6107 & $271.6$ & $25 \pm 2$ & $3^{+2}_{-2}$ & $29^{+5}_{-3}$ & $3.51 \pm 0.09$ & $38$ \\
\textbf{J1243--6423} & $297.0$ & $-19.5 \pm 0.4$ & $3.5^{+1.5}_{-0.9}$ & $3^{+2}_{-1}$ & $3.6 \pm 0.1$ & $5$ \\
J1327--6222 & $318.4$ & $-27 \pm 2$ & $2^{+2}_{-2}$ & $10^{+7}_{-4}$ & $4.7 \pm 0.2$ & $12$ \\
\textbf{J1359--6038} & $293.8$ & $2.4 \pm 0.4$ & $2.3^{+0.6}_{-0.1}$ & $2.0^{+0.2}_{-0.2}$ & $2.5 \pm 0.2$ & $2.1$ \\
J1420--6048 & $360.1$ & $73 \pm 8$ & $2^{+2}_{-2}$ & $642^{+51}_{-46}$ & $4 \pm 2$ & $669$ \\
\textbf{J1600--5044} & $262.8$ & $6.5 \pm 0.5$ & $2.3^{+1.0}_{-0.1}$ & $3.5^{+0.5}_{-0.6}$ & $2.5 \pm 0.1$ & $3.8$ \\
J1611--5209 & $127.3$ & $-10 \pm 2$ & $2^{+2}_{-2}$ & $25^{+14}_{-9}$ & $5 \pm 1$ & $25$ \\
\textbf{J1644--4559} & $478.7$ & $-26.8 \pm 0.3$ & $3.7^{+1.1}_{-0.8}$ & $5^{+3}_{-2}$ & $3.81 \pm 0.04$ & $6$ \\
J1731--4744 & $123.1$ & $-24.1 \pm 0.8$ & $4^{+2}_{-2}$ & $11^{+3}_{-2}$ & $3.6 \pm 0.1$ & $15$ \\
J1740--3015 & $151.9$ & $-12.2 \pm 0.9$ & $3^{+2}_{-1}$ & $6.9^{+1.9}_{-0.7}$ & $3.1 \pm 0.2$ & $10.0$ \\
J1824--1945 & $224.3$ & $-10.0 \pm 0.3$ & $4^{+2}_{-1}$ & $1.5^{+0.7}_{-0.4}$ & $4.26 \pm 0.08$ & $1.7$ \\
J1825--1446 & $352.2$ & $63 \pm 10$ & $2^{+2}_{-2}$ & $708^{+317}_{-223}$ & $4.6 \pm 0.8$ & $701$ \\
J1829--1751 & $216.8$ & $27 \pm 3$ & $3^{+2}_{-2}$ & $10^{+3}_{-2}$ & $3.8 \pm 0.3$ & $14$ \\
J1832--0827 & $301.2$ & $29 \pm 3$ & $2^{+2}_{-2}$ & $91^{+24}_{-19}$ & $4.2 \pm 0.2$ & $92$ \\
\textbf{J1833--0827} & $409.7$ & $-37 \pm 2$ & $2.47^{+0.09}_{-0.06}$ & $115^{+12}_{-11}$ & $2.6 \pm 0.1$ & $124$ \\
\end{longtable}

\begin{longtable}{l c c || l c c}
\caption{Pulsars with marginal multibeam DM slopes.}\\
\hline
\hline
PSR & DM & (dDM/d$t$)$_{\rm multibeam}$ & PSR & DM & (dDM/d$t$)$_{\rm multibeam}$ \\
 & [cm$^{-3}$ pc] & [$\times\, 10^{-3}$ cm$^{-3}$ pc yr$^{-1}$] &  & [cm$^{-3}$ pc] & [$\times\, 10^{-3}$ cm$^{-3}$ pc yr$^{-1}$] \\
\hline
\endfirsthead
\caption{continued}\\
\hline
PSR & DM & (dDM/d$t$)$_{\rm multibeam}$ & PSR & DM & (dDM/d$t$)$_{\rm multibeam}$ \\
 & [cm$^{-3}$ pc] & [$\times\, 10^{-3}$ cm$^{-3}$ pc yr$^{-1}$] &  & [cm$^{-3}$ pc] & [$\times\, 10^{-3}$ cm$^{-3}$ pc yr$^{-1}$] \\
\hline
\endhead
\hline
\endfoot
\hline
\noalign{\smallskip}
\multicolumn{6}{@{}p{0.8\textwidth}@{}}{\footnotesize\textbf{Notes.} Pulsars whose measured DM slopes are inconsistent with zero at the $\geq 2\sigma$ and $< 5\sigma$ level in the multibeam data. Each pulsar in the table lists the DM and the linear fit to the DM time series (dDM/d$t$)$_{\rm multibeam}$.} \\
\endlastfoot
J0034$-$0721 & $10.9$ & $-56 \pm 16$ & J1648$-$4611 & $393.2$ & $258 \pm 69$ \\
J0206$-$4028 & $13.0$ & $-71 \pm 15$ & J1650$-$4502 & $320.2$ & $-148 \pm 30$ \\
J0343$-$3000 & $20.3$ & $-393 \pm 97$ & J1651$-$4246 & $483.2$ & $86 \pm 38$ \\
J0543+2329 & $77.7$ & $4 \pm 2$ & J1652$-$2404 & $68.4$ & $-89 \pm 36$ \\
J0659+1414 & $13.9$ & $63 \pm 16$ & J1653$-$4249 & $416.1$ & $95 \pm 36$ \\
J0837$-$4135 & $147.2$ & $3.1 \pm 0.7$ & J1700$-$3312 & $166.7$ & $-220 \pm 110$ \\
J0855$-$4644 & $236.4$ & $-37 \pm 17$ & J1702$-$4310 & $377.6$ & $-77 \pm 28$ \\
J0940$-$5428 & $134.8$ & $33 \pm 10$ & J1705$-$1906 & $22.9$ & $4 \pm 1$ \\
J1016$-$5819 & $252.2$ & $-35 \pm 15$ & J1708$-$3426 & $188.7$ & $-63 \pm 22$ \\
J1020$-$6026 & $446.0$ & $691 \pm 269$ & J1716$-$4005 & $435.0$ & $-396 \pm 150$ \\
J1055$-$6028 & $636.5$ & $15 \pm 7$ & J1717$-$3425 & $583.5$ & $61 \pm 20$ \\
J1056$-$6258 & $320.7$ & $11 \pm 4$ & J1721$-$3532 & $496.8$ & $-33 \pm 14$ \\
J1112$-$6103 & $595.7$ & $-51 \pm 13$ & J1722$-$3207 & $126.1$ & $-17 \pm 7$ \\
J1123$-$6259 & $223.3$ & $68 \pm 23$ & J1726$-$3530 & $718.0$ & $-1958 \pm 578$ \\
J1136$-$5525 & $85.1$ & $-24 \pm 6$ & J1727$-$2739 & $146.0$ & $-297 \pm 78$ \\
J1138$-$6207 & $518.9$ & $-176 \pm 79$ & J1733$-$2228 & $41.1$ & $140 \pm 43$ \\
J1301$-$6305 & $374.0$ & $-292 \pm 107$ & J1735$-$3258 & $758.0$ & $1381 \pm 576$ \\
J1302$-$6350 & $146.7$ & $-19 \pm 4$ & J1739$-$2903 & $138.7$ & $18 \pm 4$ \\
J1306$-$6617 & $438.2$ & $68 \pm 34$ & J1739$-$3023 & $170.5$ & $29 \pm 12$ \\
J1317$-$6302 & $678.1$ & $-142 \pm 37$ & J1741$-$2733 & $147.4$ & $396 \pm 183$ \\
J1319$-$6056 & $400.6$ & $-43 \pm 18$ & J1741$-$3927 & $158.3$ & $46 \pm 15$ \\
J1357$-$6429 & $127.4$ & $-246 \pm 50$ & J1745$-$3040 & $88.4$ & $-2.1 \pm 0.8$ \\
J1412$-$6145 & $512.5$ & $101 \pm 41$ & J1750$-$3157 & $206.0$ & $137 \pm 63$ \\
J1430$-$6623 & $65.1$ & $-6 \pm 2$ & J1801$-$2154 & $386.0$ & $-801 \pm 188$ \\
J1435$-$5954 & $44.0$ & $-165 \pm 47$ & J1801$-$2451 & $291.6$ & $29 \pm 9$ \\
J1452$-$5851 & $260.5$ & $349 \pm 148$ & J1809$-$1917 & $197.3$ & $-33 \pm 16$ \\
J1512$-$5759 & $627.5$ & $6 \pm 2$ & J1820$-$1529 & $768.5$ & $989 \pm 236$ \\
J1522$-$5829 & $199.7$ & $-33 \pm 13$ & J1825$-$0935 & $19.4$ & $-5 \pm 2$ \\
J1524$-$5625 & $152.1$ & $-39 \pm 13$ & J1826$-$1334 & $231.7$ & $39 \pm 9$ \\
J1534$-$5334 & $25.3$ & $-31 \pm 12$ & J1830$-$1059 & $159.7$ & $-5 \pm 2$ \\
J1536$-$5433 & $147.5$ & $-529 \pm 149$ & J1834$-$0731 & $288.3$ & $-908 \pm 267$ \\
J1538$-$5551 & $602.7$ & $258 \pm 114$ & J1841$-$0425 & $325.5$ & $-10 \pm 4$ \\
J1539$-$5626 & $176.0$ & $14 \pm 3$ & J1842$-$0905 & $343.1$ & $-40 \pm 17$ \\
J1543$-$5459 & $345.0$ & $-91 \pm 20$ & J1845$-$0743 & $281.0$ & $7 \pm 2$ \\
J1602$-$5100 & $170.8$ & $6 \pm 1$ & J1847$-$0402 & $142.0$ & $12 \pm 6$ \\
J1613$-$4714 & $160.2$ & $59 \pm 17$ & J1852$-$0635 & $173.6$ & $-122 \pm 34$ \\
J1614$-$5048 & $582.4$ & $-37 \pm 13$ & J1853$-$0004 & $437.9$ & $-27 \pm 9$ \\
J1627$-$4706 & $456.1$ & $-883 \pm 354$ & J2048$-$1616 & $11.5$ & $-14 \pm 3$ \\
J1633$-$5015 & $399.0$ & $-16 \pm 6$ &  &  &  \\
\end{longtable}

\begin{longtable}{lcccccc}
\caption{Pulsars with significant UWL DM slopes.}\\
\hline
\hline
PSR & DM & (dDM/d$t$)$_{\rm UWL}$ & $\beta_{\textsc{bilby}}$ & $D_{\rm DM,\,\textsc{bilby}}(1000\,\mathrm{d})$ & $\beta_{\rm direct}$ & $D_{\rm DM,\,\rm direct}(1000\,\mathrm{d})$ \\
& [cm$^{-3}$ pc] & [$\times\, 10^{-3}$ cm$^{-3}$ pc yr$^{-1}$] & & [$\times\, 10^{-3}$ (cm$^{-3}$ pc)$^2$] & & [$\times\, 10^{-3}$ (cm$^{-3}$ pc)$^2$] \\
\hline
\endfirsthead
\caption{continued}\\
\hline
PSR & DM & (dDM/d$t$)$_{\rm UWL}$ & $\beta_{\textsc{bilby}}$ & $D_{\rm DM,\,\textsc{bilby}}(1000\,\mathrm{d})$ & $\beta_{\rm direct}$ & $D_{\rm DM,\,\rm direct}(1000\,\mathrm{d})$ \\
& [cm$^{-3}$ pc] & [$\times\, 10^{-3}$ cm$^{-3}$ pc yr$^{-1}$] & & [$\times\, 10^{-3}$ (cm$^{-3}$ pc)$^2$] & & [$\times\, 10^{-3}$ (cm$^{-3}$ pc)$^2$] \\
\hline
\endhead
\hline
\endfoot
\hline
\noalign{\smallskip}
\multicolumn{7}{@{}p{\textwidth}@{}}{\footnotesize\textbf{Notes.} Pulsars whose measured DM slopes are inconsistent with zero at the $\geq 5\sigma$ level in the UWL data. Each pulsar in the table lists the DM, the linear fit to the DM time series (dDM/d$t$)$_{\rm UWL}$, and the DM SF characterization from two independent methods: the Bayesian (\textsc{bilby}) fit and the least-squares fit. For each method, we report the power-law wavenumber spectrum index $\beta$ inferred from the DM SF and the value of the DM SF at a time lag of 1000 days, $D_{\rm DM}(1000\,\mathrm{d})$. Pulsars whose $\beta$ is constrained using the Bayesian fit are shown in bold.} \\
\endlastfoot
J0134--2937 & $21.8$ & $-1.2 \pm 0.2$ & $2^{+2}_{-2}$ & $0.05^{+0.04}_{-0.02}$ & $3.4 \pm 0.6$ & $0.06$ \\
J0525+1115 & $79.4$ & $5.0 \pm 0.6$ & $2^{+2}_{-2}$ & $0.80^{+0.07}_{-0.06}$ & $3 \pm 1$ & $0.90$ \\
J0536--7543 & $18.6$ & $6 \pm 1$ & $2^{+2}_{-2}$ & $13.1^{+0.8}_{-0.7}$ & $2 \pm 1$ & $13.8$ \\
\textbf{J0543+2329} & $77.7$ & $3.6 \pm 0.2$ & $2.28^{+0.32}_{-0.07}$ & $0.59^{+0.07}_{-0.06}$ & $2.5 \pm 0.2$ & $0.65$ \\
\textbf{J0835--4510} & $67.8$ & $-20.20 \pm 0.02$ & $3.7^{+0.2}_{-0.2}$ & $2.9^{+0.8}_{-0.7}$ & $3.69 \pm 0.04$ & $3.3$ \\
J0905--5127 & $196.4$ & $-5.8 \pm 0.8$ & $3^{+2}_{-2}$ & $0.9^{+0.2}_{-0.1}$ & $3.9 \pm 0.3$ & $1.2$ \\
\textbf{J0908--4913} & $180.2$ & $-3.50 \pm 0.03$ & $2.96^{+0.04}_{-0.04}$ & $1.9^{+0.2}_{-0.2}$ & $2.88 \pm 0.09$ & $2.1$ \\
J0940--5428 & $134.8$ & $-9 \pm 1$ & $2^{+2}_{-2}$ & $2.2^{+0.8}_{-0.6}$ & $4.7 \pm 0.3$ & $2.4$ \\
J1003--4747 & $98.3$ & $-7 \pm 1$ & $2^{+2}_{-2}$ & $1.5^{+1.0}_{-0.6}$ & $4.9 \pm 0.4$ & $1.5$ \\
\textbf{J1016--5857} & $393.6$ & $-78 \pm 3$ & $4.3^{+1.1}_{-0.9}$ & $56^{+20}_{-16}$ & $3.77 \pm 0.06$ & $68$ \\
\textbf{J1017--5621} & $437.9$ & $-4.5 \pm 0.3$ & $3.2^{+0.6}_{-0.5}$ & $0.48^{+0.11}_{-0.09}$ & $2.9 \pm 0.1$ & $0.50$ \\
J1019--5749 & $1039.9$ & $50 \pm 6$ & $2^{+2}_{-1}$ & $63^{+8}_{-4}$ & $2.3 \pm 0.2$ & $79$ \\
\textbf{J1028--5819} & $96.5$ & $-4.50 \pm 0.09$ & $3.4^{+0.2}_{-0.2}$ & $0.32^{+0.06}_{-0.06}$ & $3.15 \pm 0.04$ & $0.30$ \\
J1043--6116 & $448.8$ & $-6.2 \pm 0.6$ & $2^{+2}_{-2}$ & $0.7^{+0.1}_{-0.1}$ & $4.0 \pm 0.3$ & $0.9$ \\
J1047--6709 & $116.2$ & $3.6 \pm 0.3$ & $2^{+2}_{-2}$ & $0.81^{+0.04}_{-0.04}$ & $2.0 \pm 0.3$ & $0.90$ \\
\textbf{J1105--6107} & $271.6$ & $28.0 \pm 0.3$ & $3.6^{+0.2}_{-0.2}$ & $8^{+2}_{-2}$ & $4.05 \pm 0.08$ & $8$ \\
J1110--5637 & $261.6$ & $3.9 \pm 0.6$ & $3^{+2}_{-2}$ & $0.84^{+0.08}_{-0.06}$ & $2.6 \pm 0.3$ & $1.07$ \\
J1114--6100 & $680.3$ & $100 \pm 4$ & $4^{+2}_{-2}$ & $99^{+48}_{-29}$ & $3.5 \pm 0.1$ & $143$ \\
J1157--6224 & $324.3$ & $-11.6 \pm 0.4$ & $3^{+2}_{-2}$ & $1.7^{+0.5}_{-0.2}$ & $3.7 \pm 0.2$ & $2.7$ \\
J1210--5559 & $174.3$ & $0.58 \pm 0.08$ & $2^{+2}_{-2}$ & $0.024^{+0.002}_{-0.002}$ & $2.1 \pm 0.4$ & $0.028$ \\
\textbf{J1253--5820} & $100.6$ & $-7.9 \pm 0.3$ & $4.1^{+0.9}_{-0.7}$ & $0.6^{+0.2}_{-0.2}$ & $3.62 \pm 0.09$ & $0.6$ \\
J1320--5359 & $97.3$ & $8.0 \pm 0.9$ & $3^{+2}_{-2}$ & $1.0^{+0.2}_{-0.1}$ & $3.1 \pm 0.2$ & $1.5$ \\
J1326--6408 & $502.3$ & $5.8 \pm 0.8$ & $3^{+2}_{-2}$ & $1.2^{+0.1}_{-0.1}$ & $3.1 \pm 0.4$ & $1.5$ \\
J1326--6700 & $208.9$ & $-12.7 \pm 0.8$ & $3^{+2}_{-2}$ & $6.8^{+0.5}_{-0.3}$ & $2.6 \pm 0.3$ & $9.0$ \\
\textbf{J1327--6301} & $294.2$ & $-12.5 \pm 0.8$ & $2.5^{+1.4}_{-0.3}$ & $2.5^{+0.7}_{-0.8}$ & $2.9 \pm 0.2$ & $3.3$ \\
\textbf{J1341--6220} & $719.6$ & $-75 \pm 1$ & $3.7^{+0.9}_{-0.7}$ & $57^{+19}_{-14}$ & $3.77 \pm 0.05$ & $63$ \\
J1357--62 & $416.7$ & $-10.3 \pm 0.9$ & $2^{+2}_{-2}$ & $3.8^{+0.5}_{-0.4}$ & $3.8 \pm 0.3$ & $4.9$ \\
\textbf{J1359--6038} & $293.8$ & $18.02 \pm 0.08$ & $2.97^{+0.09}_{-0.07}$ & $3.1^{+0.6}_{-0.4}$ & $4.6 \pm 0.2$ & $2.9$ \\
\textbf{J1410--6132} & $961.0$ & $-83 \pm 2$ & $3.3^{+0.4}_{-0.4}$ & $69^{+20}_{-16}$ & $3.58 \pm 0.06$ & $76$ \\
\textbf{J1413--6141} & $667.6$ & $233 \pm 3$ & $4.0^{+0.6}_{-0.5}$ & $407^{+158}_{-118}$ & $4.08 \pm 0.02$ & $421$ \\
J1420--6048 & $360.1$ & $-13 \pm 2$ & $2^{+2}_{-2}$ & $8.0^{+0.7}_{-0.6}$ & $2.5 \pm 0.3$ & $10.0$ \\
J1424--5822 & $323.9$ & $24 \pm 2$ & $3^{+2}_{-2}$ & $6^{+1}_{-1}$ & $3.8 \pm 0.3$ & $9$ \\
\textbf{J1452--6036} & $349.5$ & $4.2 \pm 0.3$ & $3.0^{+1.1}_{-0.7}$ & $0.35^{+0.10}_{-0.09}$ & $3.0 \pm 0.1$ & $0.40$ \\
\textbf{J1453--6413} & $71.2$ & $4.83 \pm 0.08$ & $4^{+1}_{-1}$ & $0.18^{+0.10}_{-0.07}$ & $4.1 \pm 0.1$ & $0.22$ \\
\textbf{J1539--5626} & $176.0$ & $8.0 \pm 0.6$ & $2.31^{+0.31}_{-0.08}$ & $1.4^{+0.2}_{-0.2}$ & $2.6 \pm 0.2$ & $1.5$ \\
\textbf{J1557--4258} & $144.5$ & $5.7 \pm 0.3$ & $2.8^{+1.5}_{-0.6}$ & $0.4^{+0.1}_{-0.1}$ & $3.3 \pm 0.1$ & $0.5$ \\
\textbf{J1602--5100} & $170.8$ & $3.1 \pm 0.3$ & $2.27^{+0.04}_{-0.04}$ & $1.8^{+0.2}_{-0.1}$ & $2.3 \pm 0.2$ & $1.9$ \\
J1603--5657 & $264.2$ & $3.0 \pm 0.5$ & $2^{+2}_{-2}$ & $0.4^{+0.2}_{-0.1}$ & $4.7 \pm 0.6$ & $0.4$ \\
J1604--4909 & $140.7$ & $3.5 \pm 0.4$ & $3^{+2}_{-2}$ & $0.21^{+0.05}_{-0.03}$ & $3.6 \pm 0.2$ & $0.30$ \\
\textbf{J1611--5209} & $127.3$ & $-9.4 \pm 0.3$ & $4^{+1}_{-1}$ & $0.6^{+0.3}_{-0.2}$ & $4.33 \pm 0.07$ & $0.7$ \\
\textbf{J1614--5048} & $582.4$ & $25 \pm 2$ & $2.8^{+1.0}_{-0.6}$ & $12^{+2}_{-3}$ & $2.8 \pm 0.2$ & $14$ \\
J1650--4502 & $320.2$ & $-26 \pm 5$ & $2^{+2}_{-2}$ & $50^{+27}_{-17}$ & $6.6 \pm 1.0$ & $49$ \\
J1650--4921 & $228.4$ & $-12 \pm 1$ & $2^{+2}_{-2}$ & $6^{+4}_{-3}$ & $7 \pm 2$ & $6$ \\
J1701--3726 & $303.4$ & $38 \pm 4$ & $3^{+2}_{-2}$ & $58^{+6}_{-5}$ & $3.6 \pm 0.5$ & $72$ \\
J1703--3241 & $110.3$ & $-10.7 \pm 1.0$ & $2^{+2}_{-2}$ & $2.7^{+0.3}_{-0.3}$ & $3.3 \pm 0.6$ & $3.7$ \\
J1707--4053 & $351.8$ & $-39 \pm 4$ & $3^{+2}_{-2}$ & $22^{+9}_{-3}$ & $3.2 \pm 0.2$ & $35$ \\
J1709--4429 & $75.6$ & $6.3 \pm 0.3$ & $4^{+2}_{-1}$ & $0.4^{+0.2}_{-0.1}$ & $4.0 \pm 0.1$ & $0.6$ \\
\textbf{J1717--3425} & $583.5$ & $12 \pm 2$ & $2.22^{+0.06}_{-0.06}$ & $12^{+1}_{-1}$ & $2.2 \pm 0.2$ & $12$ \\
J1719--4006 & $386.1$ & $-18 \pm 2$ & $2^{+2}_{-2}$ & $5^{+2}_{-1}$ & $4.5 \pm 0.2$ & $6$ \\
\textbf{J1721--3532} & $496.8$ & $82 \pm 2$ & $3.7^{+0.9}_{-0.8}$ & $67^{+25}_{-18}$ & $3.81 \pm 0.06$ & $74$ \\
J1722--3712 & $99.4$ & $-3.0 \pm 0.5$ & $2^{+2}_{-2}$ & $0.8^{+0.4}_{-0.3}$ & $6 \pm 3$ & $0.9$ \\
J1723--3659 & $254.6$ & $-29 \pm 3$ & $2^{+2}_{-2}$ & $12^{+4}_{-2}$ & $3.8 \pm 0.3$ & $15$ \\
\textbf{J1730--3350} & $261.3$ & $-31 \pm 1$ & $3.1^{+0.3}_{-0.3}$ & $13^{+3}_{-2}$ & $3.2 \pm 0.1$ & $14$ \\
\textbf{J1731--4744} & $123.1$ & $103.0 \pm 0.2$ & $3.9^{+0.3}_{-0.3}$ & $89^{+30}_{-22}$ & $3.3 \pm 0.1$ & $110$ \\
J1733--3716 & $153.3$ & $-6.4 \pm 0.9$ & $2^{+2}_{-2}$ & $2.9^{+0.2}_{-0.2}$ & $3.3 \pm 0.7$ & $3.4$ \\
J1735--0724 & $73.5$ & $-6.5 \pm 0.6$ & $3^{+2}_{-1}$ & $0.6^{+0.3}_{-0.2}$ & $3.5 \pm 0.2$ & $0.9$ \\
\textbf{J1739--2903} & $138.7$ & $13.2 \pm 0.8$ & $2.4^{+0.5}_{-0.1}$ & $3.2^{+0.6}_{-0.5}$ & $3.1 \pm 0.2$ & $3.6$ \\
J1739--3023 & $170.5$ & $26 \pm 2$ & $3^{+2}_{-2}$ & $10^{+3}_{-2}$ & $4.4 \pm 0.2$ & $13$ \\
J1751--3323 & $295.5$ & $-56 \pm 4$ & $3^{+2}_{-2}$ & $24^{+7}_{-4}$ & $3.7 \pm 0.2$ & $40$ \\
\textbf{J1757--2421} & $179.6$ & $17.6 \pm 0.7$ & $4.0^{+0.8}_{-0.7}$ & $3.3^{+1.1}_{-0.9}$ & $3.65 \pm 0.07$ & $3.8$ \\
J1801--2304 & $1067.8$ & $-36 \pm 5$ & $3^{+2}_{-2}$ & $43^{+5}_{-4}$ & $3.5 \pm 0.4$ & $55$ \\
J1801--2451 & $291.6$ & $-15 \pm 2$ & $3^{+2}_{-2}$ & $7.9^{+1.0}_{-0.7}$ & $3.0 \pm 0.2$ & $11.0$ \\
\textbf{J1803--2137} & $234.0$ & $-27 \pm 1$ & $3.7^{+0.7}_{-0.5}$ & $10^{+3}_{-2}$ & $3.52 \pm 0.08$ & $11$ \\
J1807--0847 & $112.4$ & $-1.49 \pm 0.09$ & $2^{+2}_{-2}$ & $0.039^{+0.007}_{-0.006}$ & $4.0 \pm 0.3$ & $0.052$ \\
J1809--1917 & $197.3$ & $-27 \pm 3$ & $4^{+2}_{-2}$ & $14^{+5}_{-2}$ & $3.6 \pm 0.1$ & $21$ \\
J1817--3618 & $94.2$ & $-6.1 \pm 0.9$ & $2^{+2}_{-2}$ & $3^{+2}_{-1}$ & $5 \pm 2$ & $3$ \\
J1817--3837 & $102.7$ & $3.0 \pm 0.5$ & $2^{+2}_{-2}$ & $0.31^{+0.08}_{-0.06}$ & $3.3 \pm 0.5$ & $0.39$ \\
J1820--0427 & $84.5$ & $2.2 \pm 0.3$ & $2^{+2}_{-2}$ & $0.4^{+0.3}_{-0.2}$ & $6 \pm 2$ & $0.4$ \\
J1823--3106 & $50.3$ & $-1.9 \pm 0.2$ & $2^{+2}_{-2}$ & $0.21^{+0.14}_{-0.08}$ & $6.4 \pm 0.7$ & $0.19$ \\
\textbf{J1824--1945} & $224.3$ & $-17.78 \pm 0.05$ & $3.5^{+0.6}_{-0.5}$ & $2.2^{+1.0}_{-0.7}$ & $3.95 \pm 0.04$ & $2.6$ \\
\textbf{J1825--1446} & $352.2$ & $60 \pm 1$ & $3.4^{+1.5}_{-0.8}$ & $35^{+16}_{-12}$ & $4.8 \pm 0.2$ & $34$ \\
J1826--1334 & $231.7$ & $11 \pm 1$ & $2^{+2}_{-2}$ & $3.2^{+1.2}_{-0.9}$ & $5.2 \pm 0.5$ & $3.3$ \\
\textbf{J1828--1101} & $605.4$ & $14.1 \pm 0.9$ & $2.37^{+0.11}_{-0.07}$ & $5.1^{+0.6}_{-0.5}$ & $2.6 \pm 0.1$ & $5.4$ \\
\textbf{J1829--1751} & $216.8$ & $-18.1 \pm 0.1$ & $3.04^{+0.14}_{-0.08}$ & $4.4^{+0.7}_{-0.6}$ & $3.5 \pm 0.2$ & $4.6$ \\
\textbf{J1832--0827} & $301.2$ & $62.7 \pm 0.7$ & $3.8^{+0.4}_{-0.4}$ & $32^{+11}_{-8}$ & $3.88 \pm 0.05$ & $32$ \\
\textbf{J1833--0827} & $409.7$ & $-55.6 \pm 0.4$ & $3.24^{+0.06}_{-0.05}$ & $85^{+11}_{-10}$ & $2.77 \pm 0.09$ & $64$ \\
\textbf{J1841--0425} & $325.5$ & $11.9 \pm 0.8$ & $2.8^{+0.3}_{-0.2}$ & $5.2^{+0.9}_{-0.7}$ & $2.66 \pm 0.08$ & $6.1$ \\
J1842--0905 & $343.1$ & $-28 \pm 3$ & $3^{+2}_{-2}$ & $10^{+3}_{-2}$ & $4.1 \pm 0.2$ & $13$ \\
\textbf{J1844--0538} & $411.7$ & $31 \pm 2$ & $3.5^{+0.6}_{-0.6}$ & $13^{+4}_{-3}$ & $3.4 \pm 0.1$ & $15$ \\
J1845--0434 & $229.8$ & $23 \pm 2$ & $3^{+2}_{-2}$ & $10^{+2}_{-1}$ & $3.6 \pm 0.4$ & $14$ \\
\textbf{J1845--0743} & $281.0$ & $-2.0 \pm 0.3$ & $2.3^{+1.5}_{-0.6}$ & $0.17^{+0.05}_{-0.03}$ & $2.5 \pm 0.3$ & $0.22$ \\
J1847--0402 & $142.0$ & $-5.3 \pm 0.9$ & $2^{+2}_{-2}$ & $2.1^{+0.8}_{-0.6}$ & $4.5 \pm 0.9$ & $2.2$ \\
\textbf{J1848--0123} & $160.0$ & $-13.6 \pm 0.9$ & $2.33^{+0.06}_{-0.06}$ & $7.1^{+0.8}_{-0.7}$ & $2.6 \pm 0.2$ & $8.7$ \\
\textbf{J1853--0004} & $437.9$ & $-26 \pm 1$ & $3.0^{+0.3}_{-0.3}$ & $19^{+4}_{-3}$ & $3.0 \pm 0.2$ & $22$ \\
J1913--0440 & $89.4$ & $-1.6 \pm 0.3$ & $2^{+2}_{-2}$ & $0.23^{+0.04}_{-0.03}$ & $3.2 \pm 0.4$ & $0.28$ \\
\end{longtable}

\begin{longtable}{l c c || l c c}
\caption{Pulsars with marginal UWL DM slopes.}\\
\hline
\hline
PSR & DM & (dDM/d$t$)$_{\rm UWL}$ & PSR & DM & (dDM/d$t$)$_{\rm UWL}$ \\
 & [cm$^{-3}$ pc] & [$\times\, 10^{-3}$ cm$^{-3}$ pc yr$^{-1}$] &  & [cm$^{-3}$ pc] & [$\times\, 10^{-3}$ cm$^{-3}$ pc yr$^{-1}$] \\
\hline
\endfirsthead
\caption{continued}\\
\hline
PSR & DM & (dDM/d$t$)$_{\rm UWL}$ & PSR & DM & (dDM/d$t$)$_{\rm UWL}$ \\
 & [cm$^{-3}$ pc] & [$\times\, 10^{-3}$ cm$^{-3}$ pc yr$^{-1}$] &  & [cm$^{-3}$ pc] & [$\times\, 10^{-3}$ cm$^{-3}$ pc yr$^{-1}$] \\
\hline
\endhead
\hline
\endfoot
\hline
\noalign{\smallskip}
\multicolumn{6}{@{}p{0.8\textwidth}@{}}{\footnotesize\textbf{Notes.} Pulsars whose measured DM slopes are inconsistent with zero at the $\geq 2\sigma$ and $< 5\sigma$ level in the UWL data. Each pulsar in the table lists the DM and the linear fit to the DM time series (dDM/d$t$)$_{\rm UWL}$.} \\
\endlastfoot
J0108$-$1431 & $2.0$ & $-26 \pm 11$ & J1631$-$4721 & $873.8$ & $3390 \pm 1357$ \\
J0452$-$1759 & $39.9$ & $1.7 \pm 0.5$ & J1632$-$4621 & $562.0$ & $17 \pm 6$ \\
J0601$-$0527 & $80.5$ & $-1.3 \pm 0.6$ & J1633$-$4453 & $472.2$ & $7 \pm 2$ \\
J0627+0706 & $138.2$ & $1.3 \pm 0.5$ & J1638$-$4417 & $436.1$ & $35 \pm 11$ \\
J0631+1036 & $125.3$ & $-2.7 \pm 0.6$ & J1638$-$4608 & $423.1$ & $-19 \pm 4$ \\
J0729$-$1448 & $91.9$ & $8 \pm 2$ & J1649$-$3805 & $214.4$ & $13 \pm 3$ \\
J0758$-$1528 & $63.3$ & $-0.9 \pm 0.4$ & J1651$-$5222 & $178.3$ & $1.8 \pm 0.8$ \\
J0809$-$4753 & $228.7$ & $-2.8 \pm 0.7$ & J1653$-$3838 & $206.5$ & $2.4 \pm 0.7$ \\
J0834$-$4159 & $240.5$ & $-95 \pm 31$ & J1658$-$4958 & $192.4$ & $20 \pm 7$ \\
J0837+0610 & $12.9$ & $2.9 \pm 0.7$ & J1701$-$4533 & $526.9$ & $13 \pm 4$ \\
J0904$-$7459 & $49.2$ & $9 \pm 4$ & J1702$-$4310 & $377.6$ & $17 \pm 8$ \\
J0908$-$1739 & $15.9$ & $-1.7 \pm 0.8$ & J1705$-$3423 & $146.2$ & $3 \pm 1$ \\
J0954$-$5430 & $201.2$ & $-10 \pm 2$ & J1705$-$3950 & $207.2$ & $-5 \pm 1$ \\
J1012$-$5857 & $382.6$ & $7 \pm 1$ & J1707$-$4729 & $268.4$ & $10 \pm 2$ \\
J1015$-$5719 & $277.8$ & $-5 \pm 2$ & J1716$-$4005 & $435.0$ & $-30 \pm 6$ \\
J1046$-$5813 & $240.0$ & $-1.4 \pm 0.6$ & J1718$-$3825 & $247.5$ & $-1.7 \pm 0.8$ \\
J1055$-$6028 & $636.5$ & $-8 \pm 2$ & J1720$-$2933 & $42.7$ & $-5 \pm 2$ \\
J1057$-$5226 & $29.7$ & $0.6 \pm 0.1$ & J1722$-$3207 & $126.1$ & $1.6 \pm 0.6$ \\
J1136$-$5525 & $85.1$ & $1.5 \pm 0.6$ & J1722$-$3632 & $415.8$ & $-8 \pm 2$ \\
J1146$-$6030 & $111.7$ & $-0.7 \pm 0.3$ & J1727$-$2739 & $146.0$ & $24 \pm 5$ \\
J1156$-$5707 & $243.2$ & $-11 \pm 3$ & J1738$-$2955 & $223.4$ & $-854 \pm 337$ \\
J1301$-$6305 & $374.0$ & $-80 \pm 28$ & J1739$-$3131 & $596.9$ & $8 \pm 3$ \\
J1302$-$6350 & $146.7$ & $1.6 \pm 0.6$ & J1741$-$2733 & $147.4$ & $-28 \pm 10$ \\
J1306$-$6617 & $438.2$ & $-8 \pm 2$ & J1741$-$3016 & $382.0$ & $34 \pm 16$ \\
J1319$-$6056 & $400.6$ & $-3.7 \pm 0.9$ & J1741$-$3927 & $158.3$ & $1.9 \pm 0.8$ \\
J1435$-$5954 & $44.0$ & $-14 \pm 7$ & J1743$-$3150 & $192.2$ & $174 \pm 60$ \\
J1509$-$5850 & $142.1$ & $86 \pm 21$ & J1750$-$3157 & $206.0$ & $-17 \pm 5$ \\
J1512$-$5759 & $627.5$ & $-1.2 \pm 0.6$ & J1751$-$4657 & $20.6$ & $-1.6 \pm 0.4$ \\
J1522$-$5829 & $199.7$ & $3 \pm 1$ & J1815$-$1738 & $724.6$ & $60 \pm 29$ \\
J1530$-$5327 & $49.0$ & $5 \pm 2$ & J1834$-$0426 & $79.3$ & $-3 \pm 1$ \\
J1531$-$5610 & $110.7$ & $5 \pm 1$ & J1835$-$1106 & $132.7$ & $1.5 \pm 0.6$ \\
J1534$-$5405 & $190.6$ & $3 \pm 1$ & J1841$-$0345 & $196.6$ & $-29 \pm 6$ \\
J1544$-$5308 & $35.2$ & $0.9 \pm 0.3$ & J1843$-$0702 & $228.6$ & $-216 \pm 73$ \\
J1548$-$5607 & $314.4$ & $-7 \pm 2$ & J1852$-$2610 & $56.8$ & $7 \pm 2$ \\
J1555$-$3134 & $73.0$ & $2.0 \pm 0.9$ & J1900$-$2600 & $38.0$ & $-2.3 \pm 0.6$ \\
J1605$-$5257 & $34.8$ & $5 \pm 1$ & J1941$-$2602 & $50.1$ & $0.6 \pm 0.3$ \\
J1613$-$4714 & $160.2$ & $-4 \pm 1$ & J2330$-$2005 & $8.5$ & $-2.4 \pm 0.7$ \\
J1617$-$5055 & $467.0$ & $-43 \pm 12$ &  &  &  \\
\end{longtable}

\begin{longtable}{lcccccc}
\caption{Pulsars with significant DM slopes in the combined data.}\\
\hline
\hline
PSR & DM & (dDM/d$t$)$_{\rm UWL}$ & $\beta_{\textsc{bilby}}$ & $D_{\rm DM,\,\textsc{bilby}}(1000\,\mathrm{d})$ & $\beta_{\rm direct}$ & $D_{\rm DM,\,\rm direct}(1000\,\mathrm{d})$ \\
& [cm$^{-3}$ pc] & [$\times\, 10^{-3}$ cm$^{-3}$ pc yr$^{-1}$] & & [$\times\, 10^{-3}$ (cm$^{-3}$ pc)$^2$] & & [$\times\, 10^{-3}$ (cm$^{-3}$ pc)$^2$] \\
\hline
\endfirsthead
\caption{continued}\\
\hline
PSR & DM & (dDM/d$t$)$_{\rm UWL}$ & $\beta_{\textsc{bilby}}$ & $D_{\rm DM,\,\textsc{bilby}}(1000\,\mathrm{d})$ & $\beta_{\rm direct}$ & $D_{\rm DM,\,\rm direct}(1000\,\mathrm{d})$ \\
& [cm$^{-3}$ pc] & [$\times\, 10^{-3}$ cm$^{-3}$ pc yr$^{-1}$] & & [$\times\, 10^{-3}$ (cm$^{-3}$ pc)$^2$] & & [$\times\, 10^{-3}$ (cm$^{-3}$ pc)$^2$] \\
\hline
\endhead
\hline
\endfoot
\hline
\noalign{\smallskip}
\multicolumn{7}{@{}p{\textwidth}@{}}{\footnotesize\textbf{Notes.} Pulsars whose measured DM slopes are inconsistent with zero at the $\geq 5\sigma$ level in the UWL, multibeam, and combined data. Each pulsar in the table lists the DM, the linear fit to the DM time series dDM/d$t$, and the DM SF characterization from two independent methods: the Bayesian (\textsc{bilby}) fit and the least-squares fit. For each method, we report the power-law wavenumber spectrum index $\beta$ inferred from the DM SF and the value of the DM SF at a time lag of 1000 days, $D_{\rm DM}(1000\,\mathrm{d})$. Pulsars whose $\beta$ is constrained using the Bayesian fit are shown in bold.} \\
\endlastfoot
\textbf{J0835--4510} & $67.8$ & $-20.3 \pm 0.4$ & $3.8^{+0.3}_{-0.2}$ & $4.4^{+1.0}_{-0.8}$ & $3.56 \pm 0.08$ & $6.3$ \\
\textbf{J0908--4913} & $180.2$ & $-15.2 \pm 0.4$ & $3.5^{+0.1}_{-0.1}$ & $3.6^{+0.6}_{-0.5}$ & $3.64 \pm 0.06$ & $3.5$ \\
J1016--5857 & $393.6$ & $-59 \pm 7$ & $3^{+2}_{-2}$ & $558^{+47}_{-43}$ & $3.2 \pm 0.2$ & $619$ \\
\textbf{J1105--6107} & $271.6$ & $32 \pm 1$ & $3.9^{+0.6}_{-0.5}$ & $22^{+5}_{-4}$ & $3.39 \pm 0.09$ & $29$ \\
J1341--6220 & $719.6$ & $-59 \pm 7$ & $3^{+2}_{-2}$ & $636^{+45}_{-41}$ & $3.5 \pm 0.4$ & $693$ \\
\textbf{J1359--6038} & $293.8$ & $7.1 \pm 0.5$ & $3.1^{+0.2}_{-0.2}$ & $2.6^{+0.3}_{-0.3}$ & $3.07 \pm 0.04$ & $2.7$ \\
J1611--5209 & $127.3$ & $-12 \pm 1$ & $3^{+2}_{-2}$ & $13^{+1}_{-1}$ & $3.5 \pm 0.2$ & $15$ \\
\textbf{J1824--1945} & $224.3$ & $-13.1 \pm 0.2$ & $4.0^{+0.6}_{-0.4}$ & $2.0^{+0.6}_{-0.5}$ & $3.95 \pm 0.03$ & $2.0$ \\
J1825--1446 & $352.2$ & $76 \pm 6$ & $3^{+2}_{-2}$ & $365^{+43}_{-34}$ & $3.5 \pm 0.2$ & $424$ \\
J1832--0827 & $301.2$ & $43 \pm 2$ & $4^{+2}_{-2}$ & $62^{+12}_{-7}$ & $3.74 \pm 0.07$ & $73$ \\
\textbf{J1833--0827} & $409.7$ & $-32 \pm 3$ & $2.60^{+0.18}_{-0.10}$ & $87^{+9}_{-8}$ & $3.1 \pm 0.2$ & $88$ \\
J1739--2903 & $138.7$ & $24 \pm 3$ & $3^{+2}_{-2}$ & $71^{+7}_{-6}$ & $3.8 \pm 0.5$ & $80$ \\
J1420--6048 & $360.1$ & $60 \pm 5$ & $3^{+2}_{-2}$ & $401^{+32}_{-29}$ & $3.4 \pm 0.1$ & $451$ \\
J1845--0743 & $281.0$ & $6 \pm 1$ & $2^{+2}_{-2}$ & $12.2^{+1.0}_{-0.9}$ & $4 \pm 2$ & $12.6$ \\
\end{longtable}

\section{Extraction of the DM slope of pulsars with DM misestimation from jitter}\label{app:dm_jitter_pulsars}

For 21 slow and bright pulsars in our UWL sample, the measured DM time series exhibit significant variability that is not captured by the formal measurement uncertainties. This is primarily because of intrinsic pulse-to-pulse jitter. In these cases, the reported $1\sigma$ error bars by \textsc{tempo2} are underestimating the DM uncertainties, which can bias our linear fit.

To mitigate this and still get a first-order estimate of the DM variations, we include an Extra QUADrature (EQUAD) additional white-noise term, which adds in quadrature to the DM uncertainties. The effective DM uncertainties for each epoch are therefore
\begin{equation}
    \sigma_{\textrm{eff},\, i} = \sqrt{\sigma^2_{\textrm{DM}, \, i} + \textrm{EQUAD}^2},
\end{equation}
where $\sigma^2_{\textrm{DM}, \, i}$ are the formal uncertainties per epoch quoted by \textsc{tempo2}. The DM slope is then estimated using a weighted linear least-squares fit to the time series using these effective uncertainties. Using this procedure, we find three pulsars that show significant DM variations with these inflated uncertainties. 

\begin{figure*}[ht!]
    \centering
    \begin{subfigure}[t]{0.33\textwidth}
        \includegraphics[width=\textwidth]{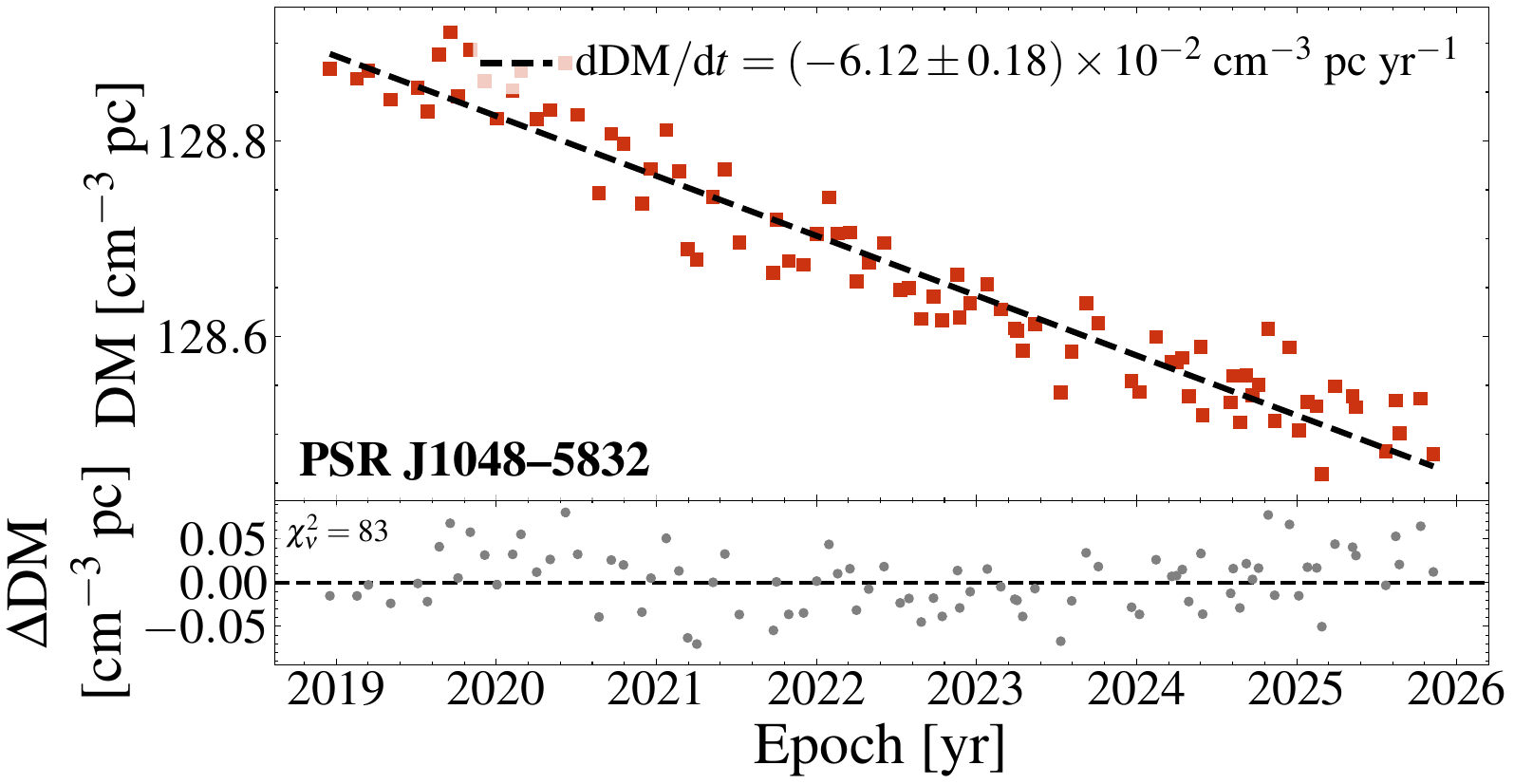}
    \end{subfigure}
    \hfill
    \begin{subfigure}[t]{0.33\textwidth}
        \includegraphics[width=\textwidth]{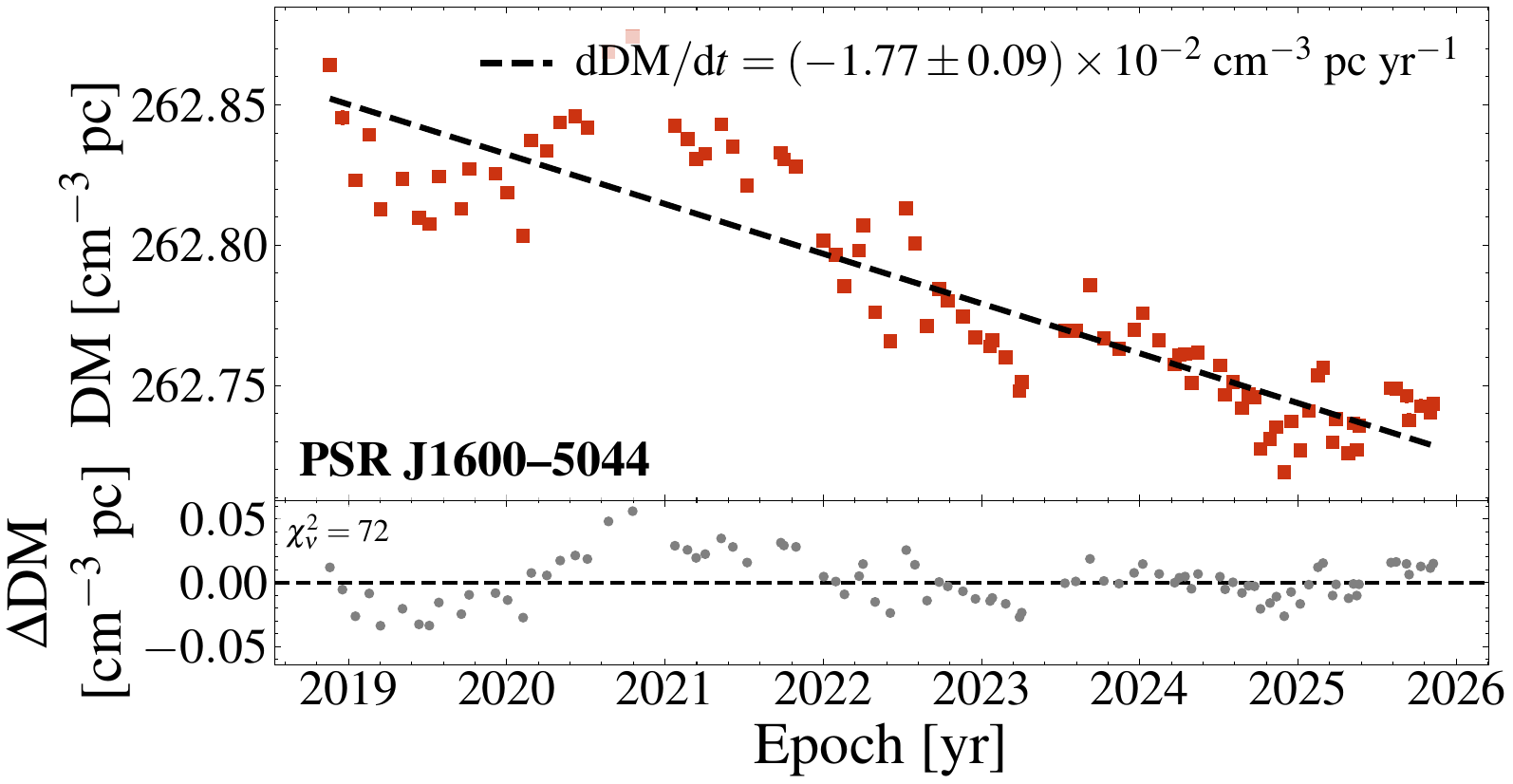}
    \end{subfigure}
    \hfill
    \begin{subfigure}[t]{0.33\textwidth}
        \includegraphics[width=\textwidth]{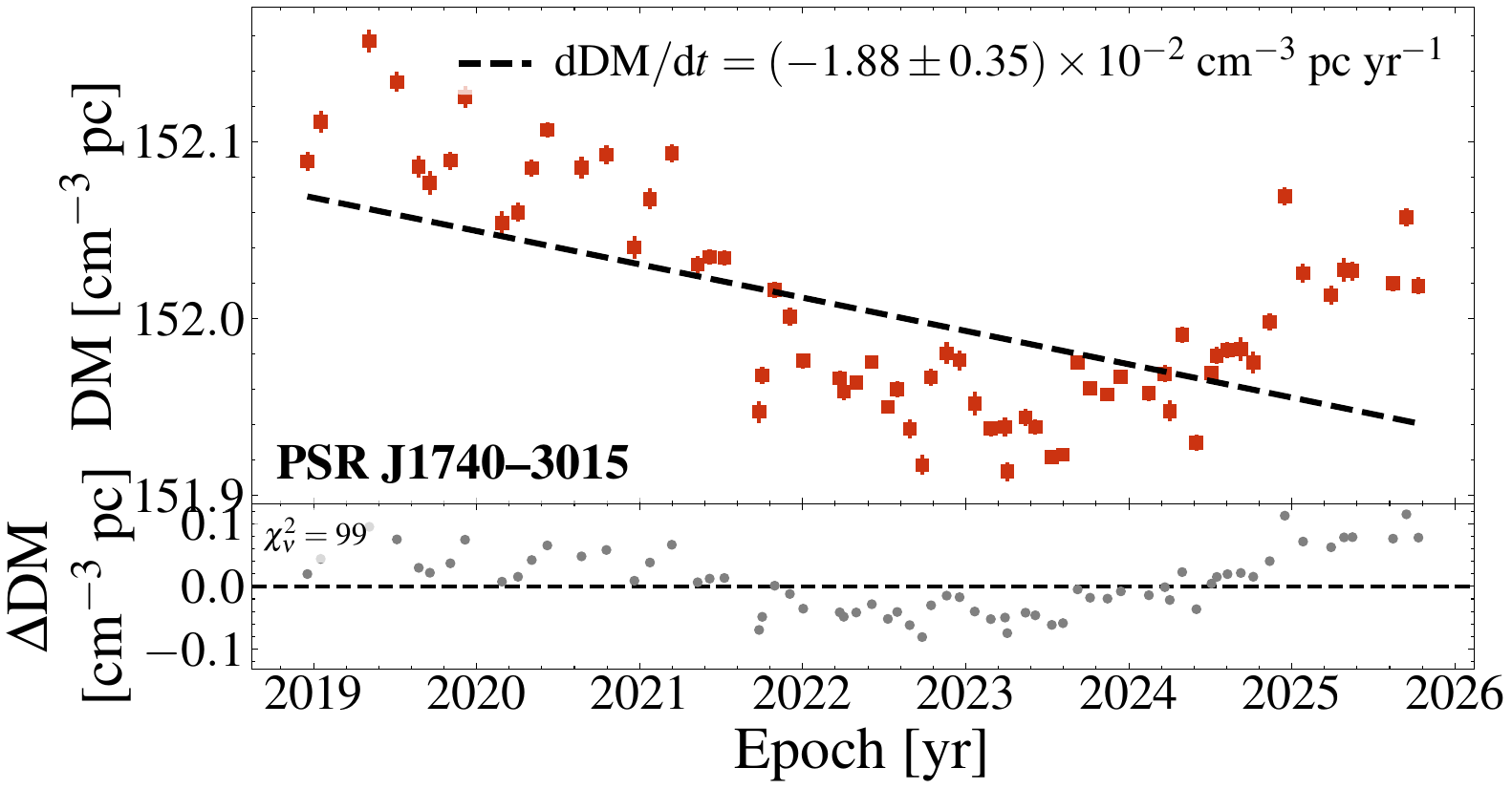}
    \end{subfigure}
    \begin{subfigure}[t]{0.33\textwidth}
        \includegraphics[width=\textwidth]{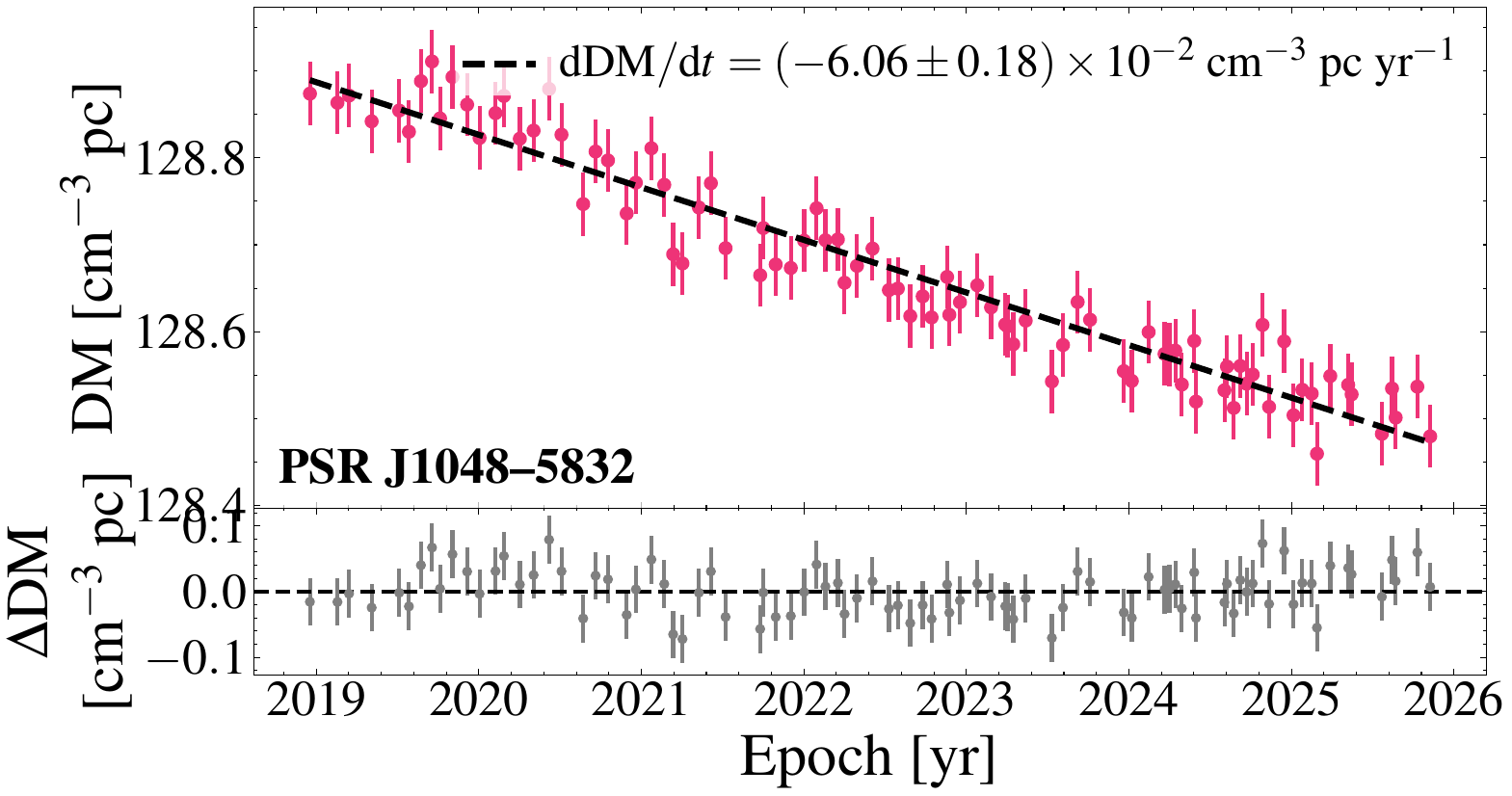}
    \end{subfigure}
    \hfill
    \begin{subfigure}[t]{0.33\textwidth}
        \includegraphics[width=\textwidth]{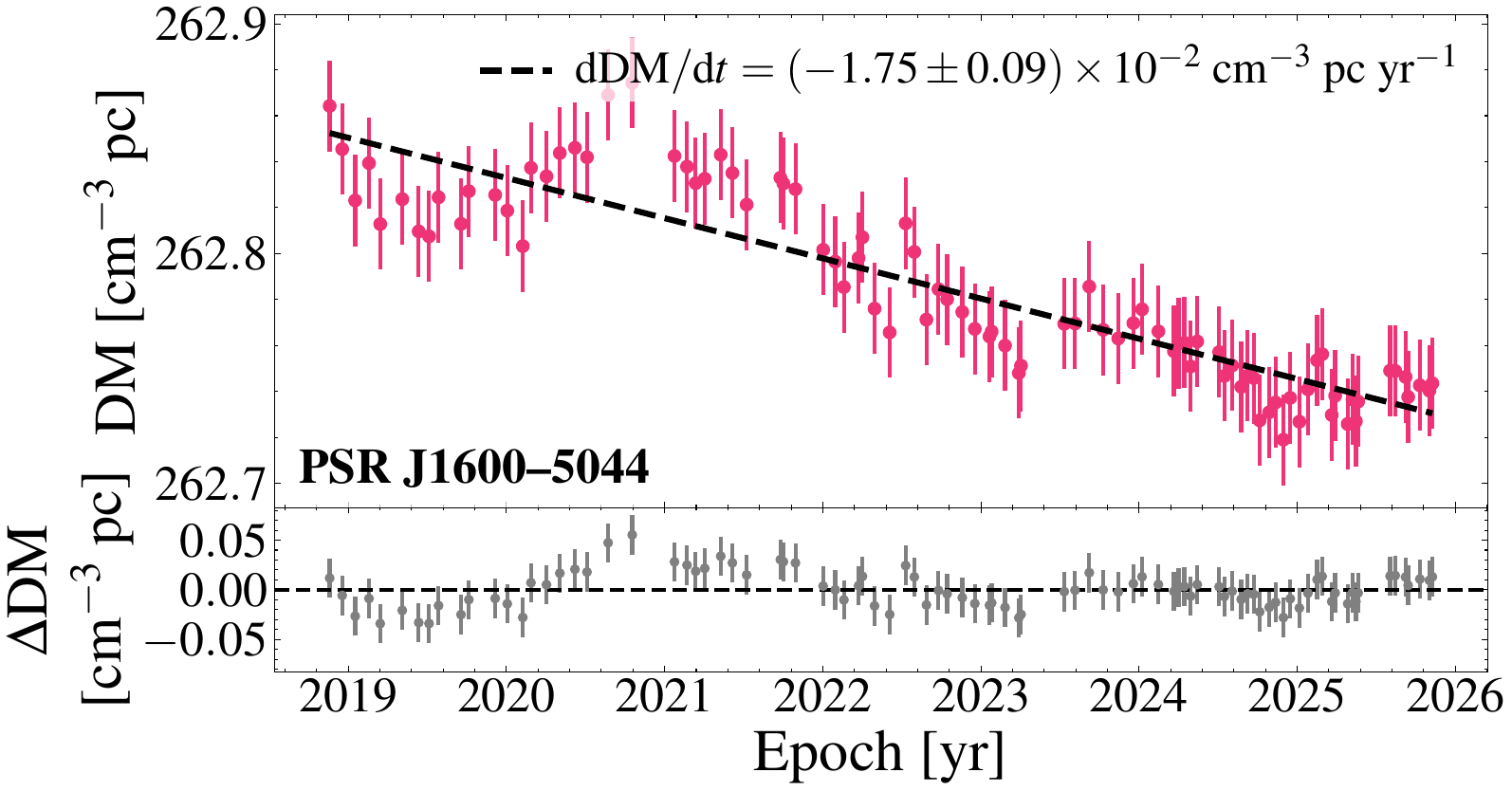}
    \end{subfigure}
    \hfill
    \begin{subfigure}[t]{0.33\textwidth}
        \includegraphics[width=\textwidth]{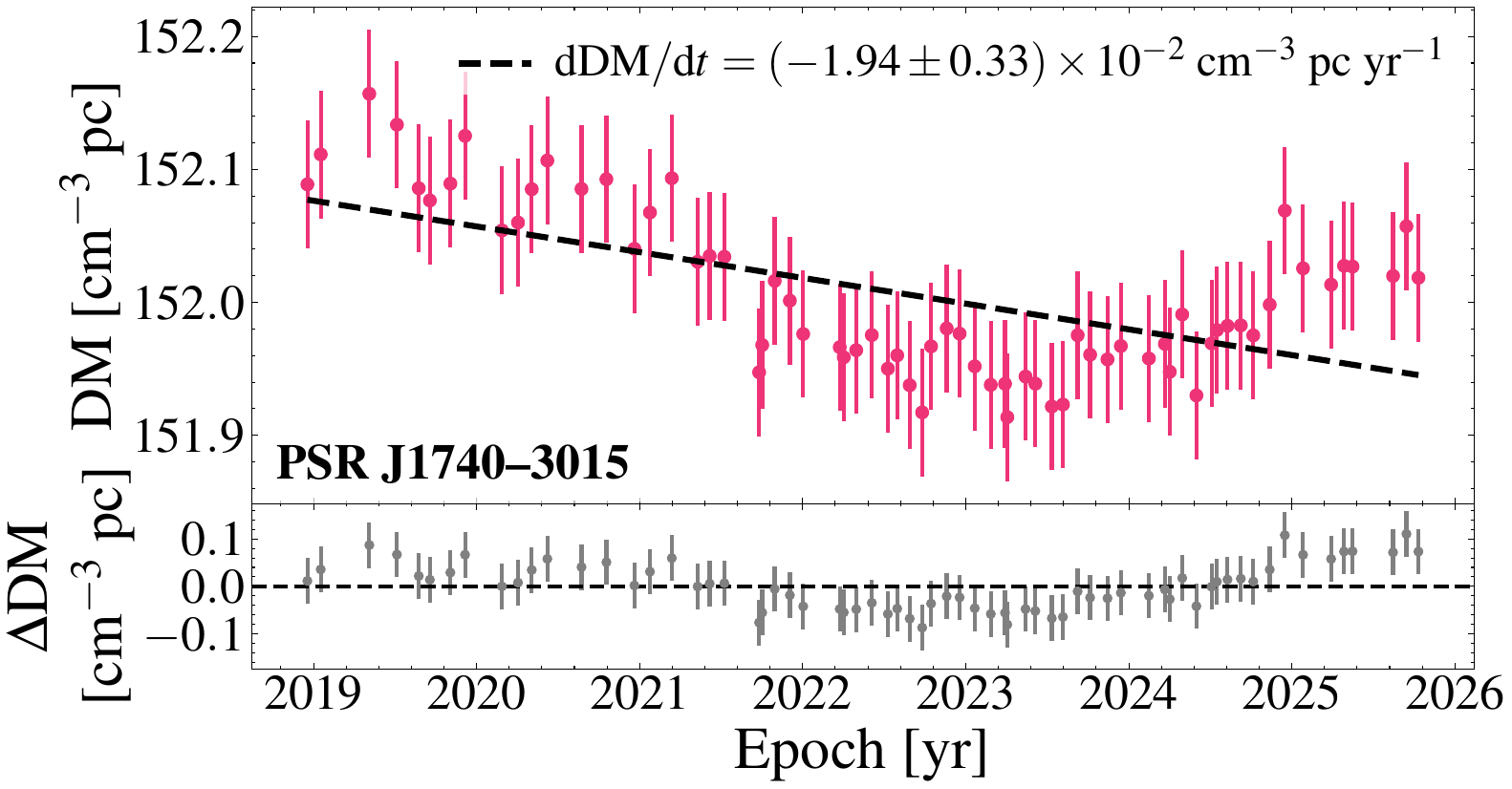}
    \end{subfigure}
    \caption{\textit{Top}: DM time series for PSRs~J1048--5832 (\textit{left}), J1600--5044 (\textit{middle}), and J1740--3015 (\textit{right}). The red squares show the measured DM values with their corresponding $1\sigma$ uncertainties for the UWL dataset. The reduced chi-squared ($\chi_\nu^2$) was computed after removing the linear trend. \textit{Bottom}: Same as the top but now the effective DM uncertainties in the pink points are used by adding an EQUAD. The gray points show the DM values after subtracting the fitted linear trend. }
\label{fig:pulsars_jitter}
\end{figure*}

Figure~\ref{fig:pulsars_jitter} shows the DM time series of these three pulsars before and after adding EQUAD. In Table~\ref{table:jitter_results} we report the ratio between EQUAD and the median DM uncertainties. High values of this ratio indicate that the DM time series are dominated by the pulse-jitter rather than the radiometer noise. 

\begin{table}[ht!] 
\caption{Pulsars with significant UWL DM slopes after fitting an EQUAD parameter.}              
\label{table:jitter_results} 
\centering
\begin{tabular}{lccccc}
\hline\hline

PSR & DM & EQUAD / med($\sigma_{\rm DM}$) & (dDM/d$t$)$_{\rm UWL}$ \\
& [cm$^{-3}$ pc] & & [$\times\, 10^{-3}$ cm$^{-3}$ pc yr$^{-1}$] \\
\hline\hline               
J1048$-$5832 & 128.7 & 10.0 & $-61 \pm 2$ \\
J1600$-$5044 & 262.9 & 9.3 & $-18 \pm 1$  \\
J1740$-$3015 & 151.8 & 9.5 & $-19 \pm 3$  \\
\hline
\end{tabular}
\tablefoot{Pulsars whose measured DM slopes are inconsistent with zero at the $\geq 5\sigma$ level in the UWL dataset after fitting an EQUAD to the DM error bars. The table lists the DM, the ratio between EQUAD and the median of the DM error, and the linear fit to the DM time series after adding EQUAD to the DM uncertainties.}
\end{table}

\end{appendix}

\end{document}